\documentclass[a4paper,manyauthors,nocleardouble,COMPASS]{cernphprep}

\RequirePackage[T1]{fontenc}

\RequirePackage{graphicx}
\RequirePackage{newtxtext}
\RequirePackage{newtxmath}
\RequirePackage{flushend}
\RequirePackage[numbers,sort&compress]{natbib}

\usepackage{etex}
\usepackage{amsmath,amssymb,epsfig,fleqn,url,color,here,graphics,graphicx,rotating,multirow,wrapfig}
\usepackage{pbox}
\usepackage{dcolumn}
\usepackage{bigstrut}
\usepackage{soul,xcolor}
\usepackage{lineno, xcolor}

\usepackage[section]{placeins}
\usepackage{orcidlink}
\usepackage{tikz}
\usepackage{hyperref}
\hypersetup{
    colorlinks,
    citecolor=blue,
    filecolor=blue,
    linkcolor=blue,
    urlcolor=blue
}

\setstcolor{red}
\begin{document}


\begin{titlepage}

\PHnumber{2024-xxx}
\PHdate{\today}
\title{Multiplicities of positive and negative pions, kaons and unidentified hadrons from deep-inelastic scattering of muons off a liquid hydrogen target}

\date{}
\Collaboration{The COMPASS Collaboration}
\ShortAuthor{The COMPASS Collaboration}

\begin{abstract} \label{abstract}

The multiplicities of positive and negative pions, kaons and unidentified hadrons produced
in deep-inelastic scattering are measured in bins of the Bjorken scaling variable $x$,
the relative virtual-photon energy $y$ and the fraction of the
virtual-photon energy transferred to the final-state hadron $z$.
Data were obtained by the COMPASS Collaboration using a 160 GeV muon beam of both electric charges
and a liquid hydrogen target. These measurements cover the kinematic domain with photon virtuality $Q^2 > 1$ (GeV/$c)^2$, $0.004 < x < 0.4$, $0.1 < y < 0.7$ and $0.2 < z < 0.85$, in accordance with the kinematic domain used in earlier published COMPASS multiplicity measurements with an isoscalar target. The calculation of radiative corrections was improved by using the Monte Carlo generator DJANGOH, which results in up to 12\% larger corrections in the low-$x$ region.

\end{abstract}

\vspace{2cm}

\centerline{\it This article is dedicated to the memory of Roland Windmolders}

\vfill
\Submitted{(to be submitted to PRD)}

\end{titlepage}


\maketitle

\section{Introduction}

The measurement of hadron production in deep-inelastic lepton--nucleon
scattering (SIDIS), $\ell$\,N $\rightarrow$ $\ell$\,h\,X, is an important method for studying the structure and formation of
hadrons. In the framework of perturbative quantum chromodynamics (pQCD), the process of parton
fragmentation into hadrons is described by non-perturbative fragmentation functions (FFs), which are crucial in understanding the transition from quarks to
observable hadrons.
As FFs cannot yet be predicted by theory, they must presently be determined through measurements.
They are believed to be universal across various high-energy processes.
Studying them in different experiments and exploring various kinematic regions leads to a more comprehensive understanding of hadron formation in the context of pQCD.

In order to study FFs, the cleanest approach is to investigate
hadron production in annihilation processes,
such as e$^+$e$^-\rightarrow$ h\,X.
 High-precision data were collected by various experiments,
 including ALEPH, DELPHI and OPAL at LEP~\cite{lep}, SLD and BABAR at SLAC~\cite{slac}
 and BELLE at KEK~\cite{belle}. These experiments cover a wide range of centre-of-mass energies
from 10 GeV up to the Z$^0$-mass.
However, e$^+$e$^-$ annihilation
primarily provides information about
the sum of quark and anti-quark FFs, and only limited flavour separation is possible unless
model-dependent algorithms for quark-flavour tagging are employed.
In contrast, in SIDIS analyses it is possible
to access and separate quark and anti-quark contributions.
Here, FFs appear convoluted with
parton distribution functions (PDFs)
in the pQCD description of
the measurement, thus full flavour separation is in principle possible.
 Most SIDIS data were collected in fixed-target experiments, such as
HERMES at DESY \cite{hermes}, and EMC and COMPASS at CERN \cite{emc, comp_pi, comp_K}. Due to the limited range in centre-of-mass energy in these experiments,
the ability to access gluon FFs through the study of scaling violations is limited.
The gluon FFs are indirectly probed by hadron--hadron collisions, $e.g.$
at RHIC \cite{rhic}, using for example
single-inclusive hadron production at high transverse momentum
\cite{dss_01}.
All these data sets have been analysed by several theoretical and phenomenological groups to obtain FFs \cite{hkns, lss, dss_02, dss_03, nnpdf_ff,jam_ff,nnlo_ff}.

This paper presents the results of COMPASS measurements on
multiplicities of positvie and negative  pions, kaons and unidentified hadrons produced in muon-proton interactions.
These measurements complement our earlier results on multiplicities obtained using an isoscalar target. 
The available centre-of-mass energy allows COMPASS to cover a kinematic range larger than that of HERMES  \cite{hermes} and similar to that of EMC \cite{emc}.

\section{Formalism}

The multiplicity for a hadron of type h measured in SIDIS is defined as
the differential hadron
production cross section $\sigma^{\rm h}$ normalised to the inclusive deep inelastic
scattering (DIS) cross section $\sigma^{\rm DIS}$:

\begin{equation} \label{eq:1}
    \frac{{\rm d}M^{\rm h}(x,Q^2,z)}{{\rm d}z} =
    \frac{{\rm d}^3\sigma^{\rm h}(x,Q^2,z)/{\rm d}x{\rm d}Q^2 {\rm d} z}
    {{\rm d}^2\sigma^{\rm DIS}(x,Q^2)/{\rm d}x {\rm d} Q^2}.
\end{equation}

Here, $x$ denotes the Bjorken scaling variable, $Q^2$ the virtuality of the photon
mediating the lepton--proton scattering process and $z$
the fraction of the virtual-photon energy carried by the
produced hadron in the target rest frame.
When these multiplicities are integrated over the variable $z$, they represent the average number of hadrons of type h produced per DIS event.
Other variables used
are the lepton energy fraction carried by the virtual-photon $y$, the energy of the virtual-photon in the laboratory frame $\nu$
and the invariant mass of the final hadronic system $W$. For Lorentz-invariant definitions of the above variables see $e.g.$ Ref.~\cite{comp_pi}.

Within the factorisation framework of pQCD, $\sigma^{\rm DIS}$ can be expressed through a sum over parton flavours. For each patron type $a=$\{q, $\bar{\rm q}$, g\} the corresponding PDF is convoluted with the lepton--parton hard-scattering cross section. In the cross section expression for the hadron production in the current fragmentation region, $\sigma^{\rm h}$, the sum over parton flavours involves
additional convolutions of these PDFs with fragmentation functions. In leading order (LO) the differential cross sections
$\sigma^{\rm h}$ and $\sigma^{\rm DIS}$ can be expressed as

\begin{equation}
 \frac{ {\rm d}^3 \sigma^{\rm h}(x,Q^2,z) }{{ \rm d} x{\rm d} Q^2 {\rm d} z }   =
 \frac{2 \pi \alpha_{\rm em}^2 (1+(1-y)^2)}{Q^4}
 \sum_{a} e_a^2 \, q_a(x,Q^2) D_a^{\rm h}(Q^2,z),
 \end{equation} \label{eq:2}

\begin{equation}
\frac{{\rm d}^2\sigma^{\rm DIS}(x,Q^2) }{ {\rm d} x {\rm d} Q^2} =  \frac{2 \pi \alpha_{\rm em}^2 (1+(1-y)^2)}{Q^4} \sum_{a} e_a^2 \, q_a(x,Q^2).
\end{equation} \label{eq:3}

Here, $\alpha_{\rm em}$ is the fine structure constant,
$e_a$ the fractional electric charge of quarks of species $a$,
 $q_a(x, Q^2)$ the quark PDF for the flavour $a$, and $D_a^{\rm h}(Q^2, z)$ the fragmentation function
of the quark of flavour $a$ to the hadron of type h.
In LO pQCD, FFs have a probability interpretation
similar to PDFs,
namely $D_a^{\rm h}$ denotes the number density of
hadrons h produced in the hadronisation of quarks of species $a$. The fragmentation of a quark of a given species into a final-state hadron is called favoured, $D_{\rm fav}$, if the quark
flavour corresponds to a valence quark in the hadron, otherwise the fragmentation is called unfavoured $D_{\rm unf}$.
The more complicated NLO pQCD formulas for the cross sections can be found $e.g.$
in Ref.~\cite{nlo_sidis}.

\section{Experimental setup and data analysis}

The data were collected in 2016 using muons of both electric charges from the M2 beamline at the CERN SPS. The beam had a momentum of 160~GeV/$c$ with a variation of  $\pm$ 5\%
and a typical root mean square size at the target position of $7 \times 7$~mm$^2$.
The beam was naturally polarised, but the polarisation is not affecting this analysis since we integrate over azimuthal angle and transverse momentum of the produced hadrons, and we also sum the results of both beam charges.
The beam was delivered in cycles typically of 36~s, consisting of two spills,
each lasting 4.8~s. The intensity of the beam was $1.6 \times 10^7$~s$^{-1}$,
which is a factor of three lower than that of previous COMPASS measurements on the isoscalar target \cite{comp_pi,comp_K}, leading to lower
data statistics but a more stable spectrometer.

The incident muons were impinging on a liquid hydrogen target with a total length of 250~cm
and a diameter of approximately 4~cm. The 2016 configuration of the COMPASS spectrometer was
optimised for the precise measurement of exclusive processes, such as deeply virtual Compton
scattering. The target was surrounded by
a time-of-flight detector,
designed to measure and identify recoil protons.
While, in principle, this detector has the potential to assist in the
exclusion of diffractive events, its data are not used in the present analysis.

The two-stage COMPASS spectrometer has been designed to reconstruct scattered muons and produced
hadrons in a wide range of polar and azimuthal angles and momentum.
For particle tracking various detectors surrounding the two spectrometer magnets are employed. The direction of the reconstructed tracks at the
interaction point is determined with a precision of 0.2~mrad. The momentum resolution is about 1.2\% in the first spectrometer stage and is further improved to 0.5\% in the second stage.
The identification of muons is performed using hadron absorbers.

In the first spectrometer stage, a Ring Imaging Cherenkov counter (RICH) is employed for the separation of pions, kaons and protons \cite{comp_exp}.
This detector uses C$_{4}$F$_{10}$ as the radiator gas
corresponding to momentum thresholds of approximately 2.9~GeV/$c$ for pions, 9~GeV/$c$ for kaons and 18~GeV/$c$ for protons.
Within the central region of the RICH, photon detection is accomplished using multi-anode photomultiplier tubes known for their high photodetection efficiency and rapid response,
ideal for operation in a high-rate environment.

The trigger system, based on four pairs of scintillator hodoscopes, was designed to select scattered muons with
a minimum scattering angle.
In contrast to earlier measurements with an isoscalar target, an additional trigger is employed covering the low-$x$ and high-$y$ domain.
The existing trigger designed for the large-$x$ and large-$Q^2$ region is not used in the present analysis.

The data analysis includes several steps: event selection, particle identification, corrections for spectrometer acceptance, as well as corrections for QED radiative effects and for contamination by diffractively produced vector-mesons.
The multiplicities denoted as $M^{i}(x, y, z)$, for hadrons of type $i =\{\pi^{+}$, $\pi^{-}$, $\rm{K}^+$, $\rm{K}^-$, $\rm{h}^+$, $\rm{h}^-\}$, where h$^{+}$, h$^{-}$ stand for unidentified hadrons, is given by:
\begin{equation} \label{eq:mult_data}
    \frac{dM^{i}(x,y,z)}{dz} = \frac{1}{N^{\rm DIS}(x,y)} \frac{dN^i(x,y,z)}{dz} \frac{1}{A^{i}(x,y,z)}.
\end{equation}
Here, $N^{\rm DIS}$ denotes the number of DIS events, while $N^i$ and $A^i$ denote the yield of particles of type $i$ and the respective acceptance correction factor.
In the present analysis, similar as in earlier determinations of multiplicities using COMPASS data taken with an isoscalar target, $y$ is used as the third variable in order to cope with the significant correlation between $x$ and $Q^2$ inherent in fixed-target measurements.

\subsection{Event and hadron selection}

\begin{figure}
    \centerline{
    \includegraphics[trim=0cm 0cm 0.8cm 0cm,clip,width=0.33\textwidth]{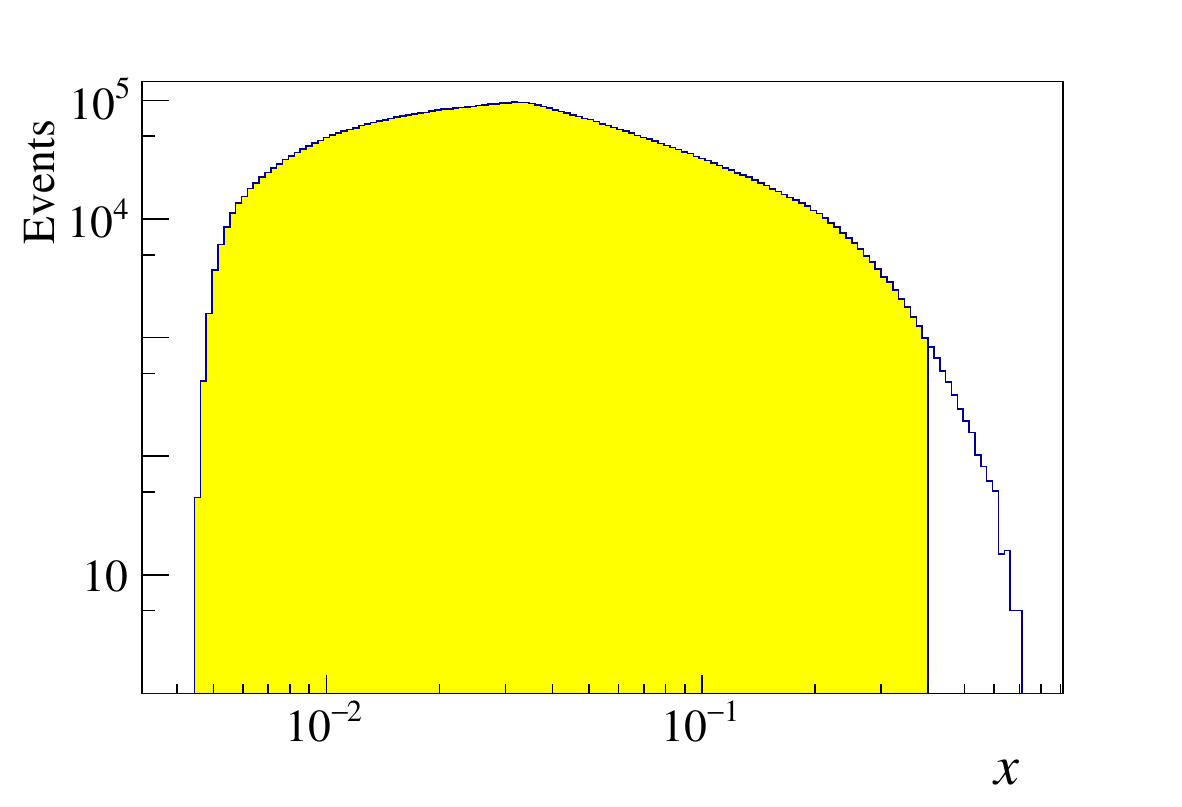}
    \includegraphics[trim=0cm 0cm 0.8cm 0cm,clip,width=0.33\textwidth]{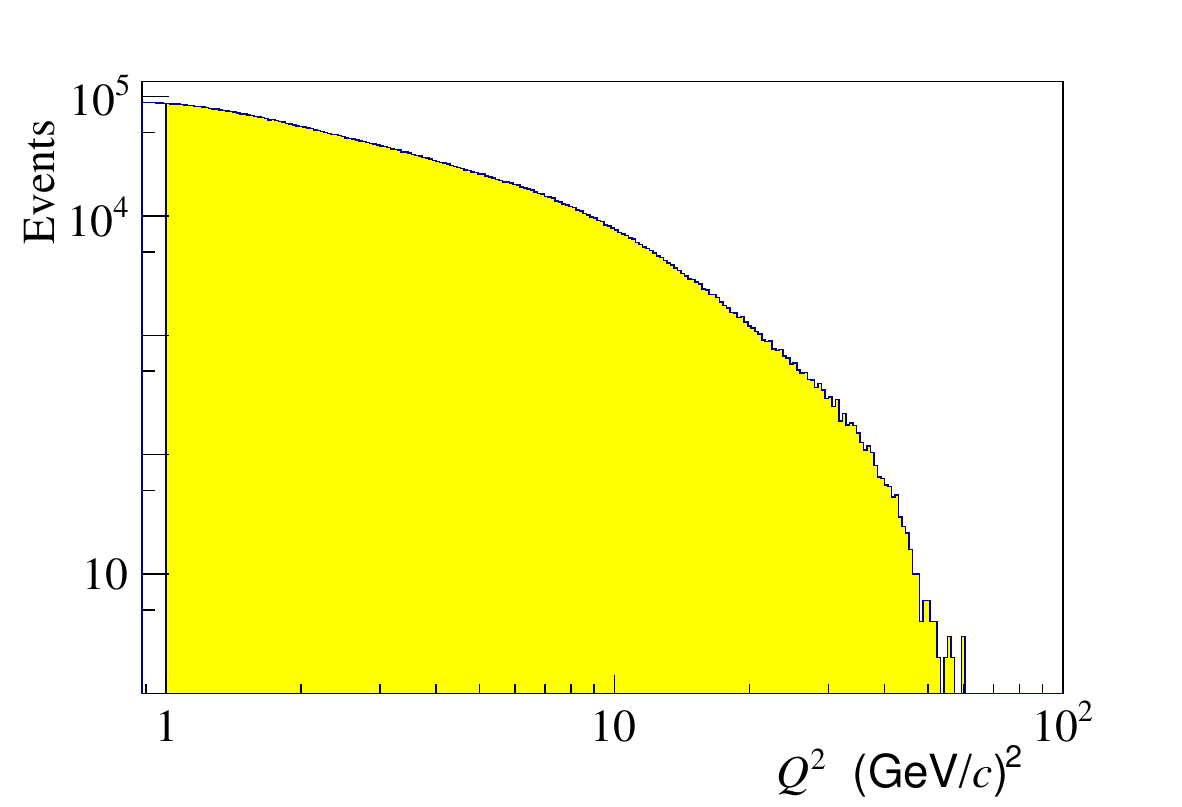}
    \includegraphics[trim=0cm 0cm 0.8cm 0cm,clip,width=0.33\textwidth]{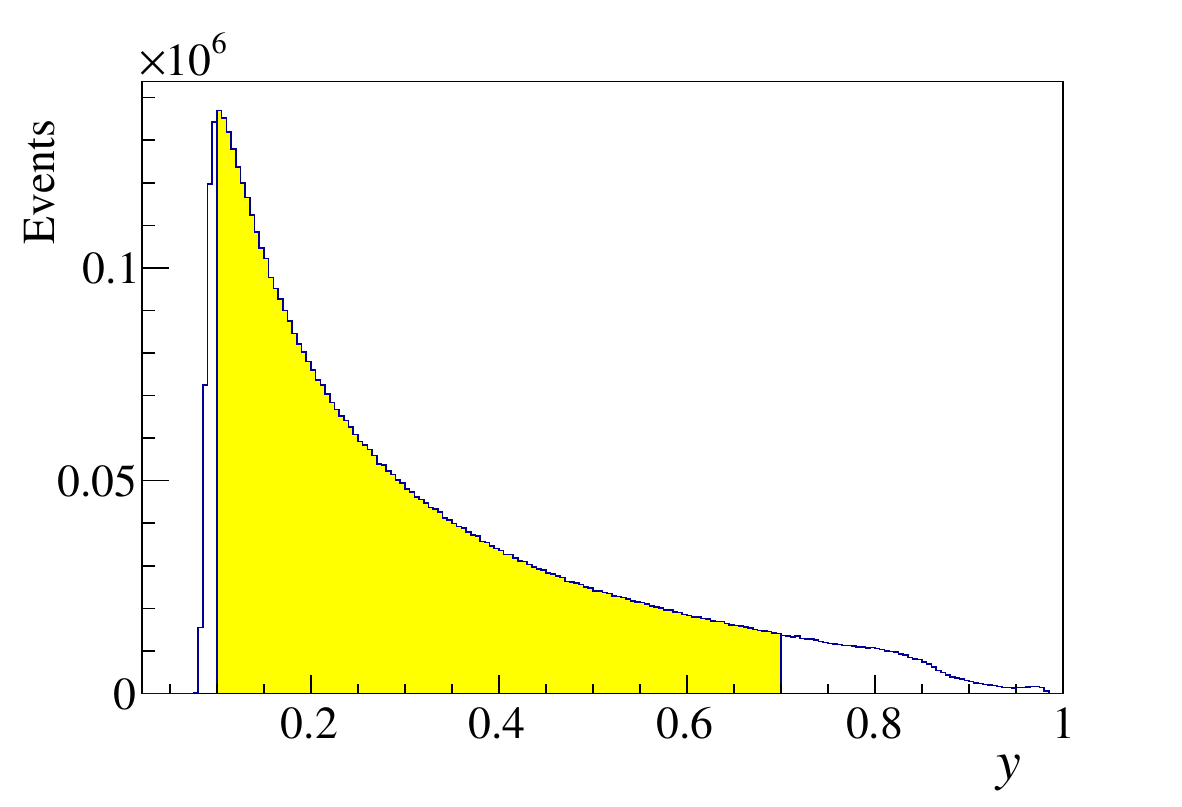}}
    \centerline{
    \includegraphics[trim=0cm 0cm 0.8cm 0cm,clip,width=0.33\textwidth]{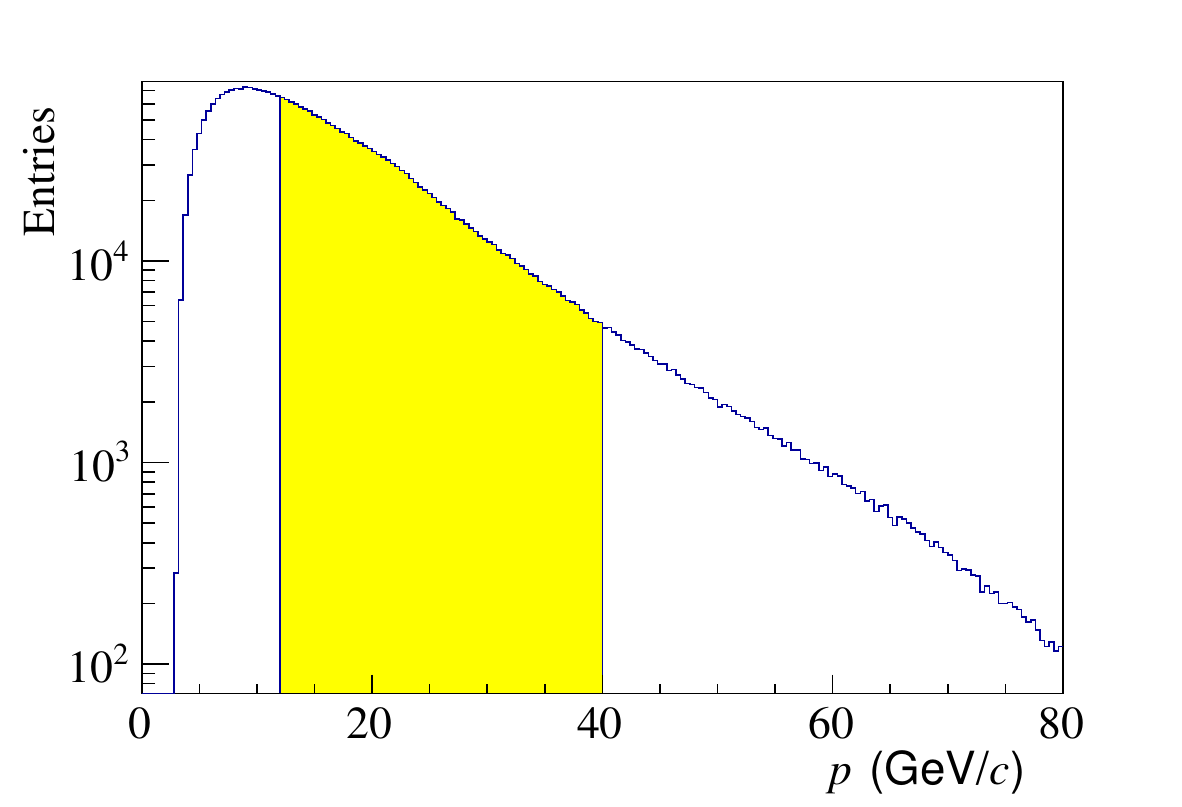}
    \includegraphics[trim=0cm 0cm 0.8cm 0cm,clip,width=0.33\textwidth]{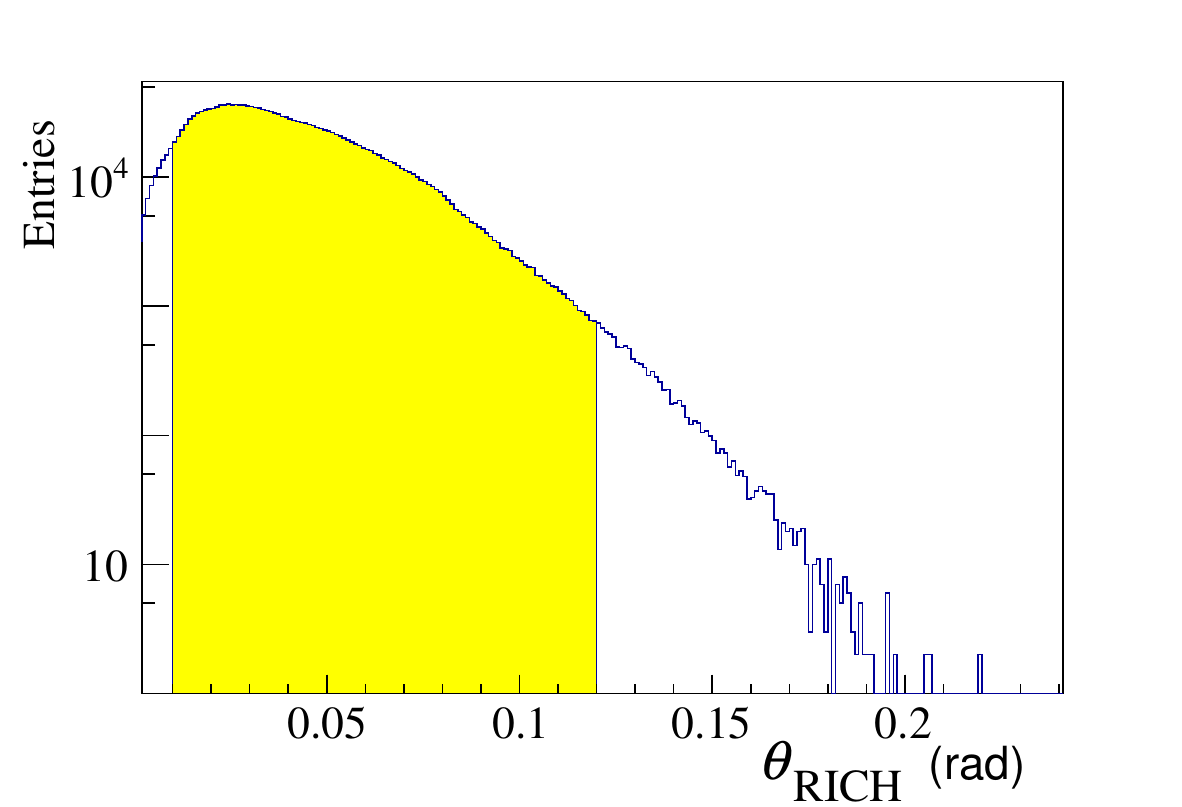}
    \includegraphics[trim=0cm 0cm 0.8cm 0cm,clip,width=0.33\textwidth]{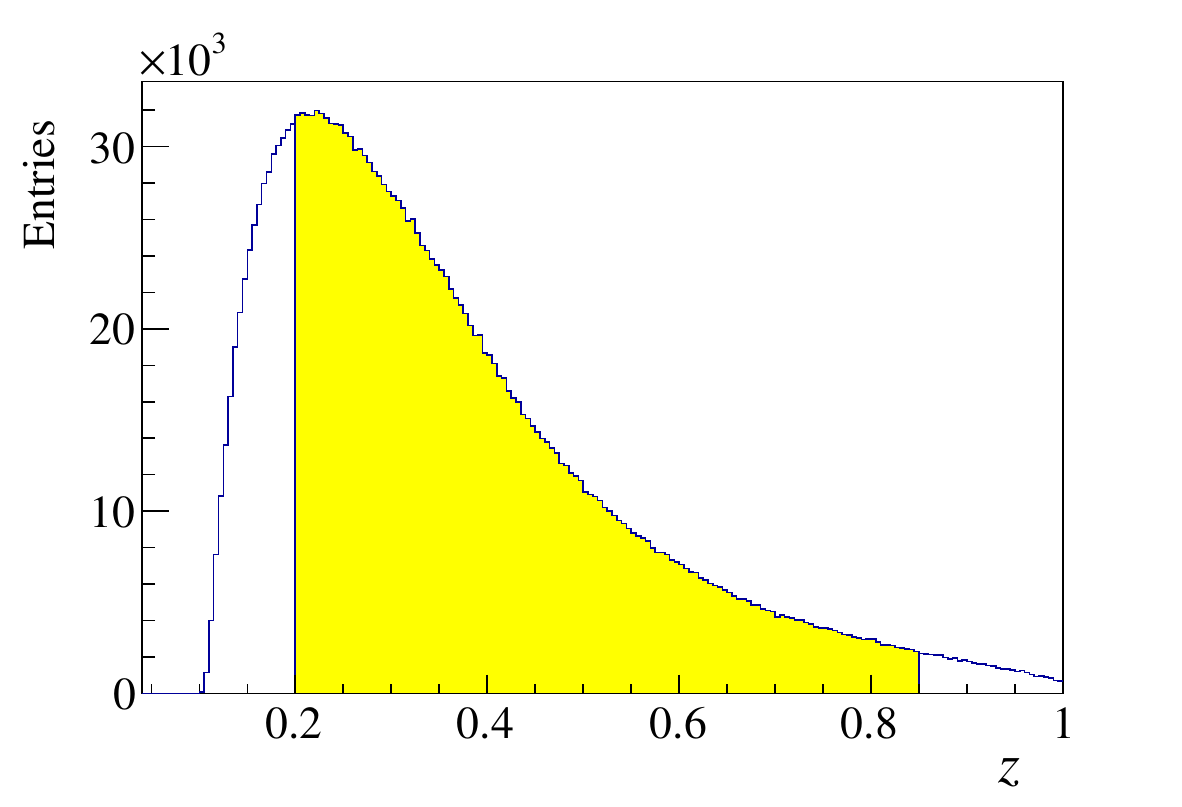}}

    \caption{Top panels: inclusive variables $x$, $Q^2$ and $y$; bottom panels: hadron variables $p$, $\theta_\textup{RICH}$ and $z$. The selected regions are marked as filled areas.}
    \label{fig:kine}
\end{figure}

Consistent with previous analyses,
the present study is based on events selected by triggers that use information related to scattered muons only. In order to be accepted, events are required
to have a reconstructed interaction vertex associated with both an incoming and a scattered muon track and with a vertex positioned within the defined fiducial target volume.
Furthermore, the incident muon momentum, which is measured by a dedicated set of scintillating fibre detectors
upstream of the COMPASS spectrometer,
is constrained to the interval between 140 GeV/$c$ and 180 GeV/$c$.
Events are accepted if $Q^2 > 1$ (GeV/$c)^2$, $0.004 < x <0.4$ and $W > 5$ GeV/$c^2$. These requirements select the deep-inelastic scattering regime, avoiding the
nucleon resonance region.
The relative virtual-photon energy is limited to the interval $0.1 < y < 0.7$.
The lower bound is used to exclude kinematic regions where the momentum resolution deteriorates, and the upper bound removes the kinematic range where radiative effects become
particularly pronounced. The analysis comprises 5.4~million selected inclusive
DIS events.
For these events, the top panels of Fig.~\ref{fig:kine} show distributions and applied constraints in $x$, $Q^2$ and $y$, which are hereafter referred to as 'inclusive variables'.

For a selected DIS event, a reconstructed track is considered to be a hadron, if it originates from the interaction vertex and does not cross muon filters.
The fraction of the virtual-photon energy transferred to a final-state hadron is restricted to
$0.2 < z <0.85$, where for an unidentified charged hadron the pion mass is assumed.
The lower limit avoids contamination from target remnant fragmentation,
while the upper one excludes muons wrongly identified as hadrons.
For pion and unidentified hadrons the upper limit also
excludes the region with large diffractive contributions, while for kaons it removes the domain where pQCD might not be applicable as suggested by previous COMPASS analyses \cite{comp_rk, comp_rkp}.

In order to take into account the acceptance and performance of the RICH detector, a hadron is accepted only if it has a momentum between $p_{\rm min}=12$~GeV/$c$ and
$p_{\rm max}= 40$~GeV/$c$, and a polar angle in the range $0.01~<~\theta_{\rm RICH}~<~0.12$ with a vertical projection $\theta_{\rm RICH}^{Y} < 0.08$.
Moreover, to only consider $(x, y)$ regions where hadron detection is possible,
 the following constraint is imposed bin-by-bin:
\begin{equation*}
    \frac{\sqrt{p_{\min}^2c^2 + m_i^2c^4}}{z_{\min}} < \nu < \frac{\sqrt{p_{\max}^2 c^2 + m_i^2 c^4}}{z_{\max}}.
\end{equation*}
Here, $z_{\min}$ and $z_{\max}$ correspond
to the boundaries of a given bin in the $z$ variable.
It is worth mentioning that these restrictions on $\nu$ are crucial,
as they serve to reduce a possible dependence of the analysis results on the physics model embedded within the LEPTO generator, which is used for acceptance calculations. The same restrictions
were already used in previous COMPASS analyses \cite{comp_pi, comp_K}.

The bottom panels of Fig.~\ref{fig:kine} show distributions and applied constraints in $p$,  $\theta_\textup{RICH}$ and $z$, which are hereafter referred to as 'hadron variables'. The entire set of selected events comprises 1.7 million unidentified hadrons, 1.3 million pions and 280 thousand kaons.
The amount of kaons in this dataset is about 50\%
of the one used in Ref.~\cite{comp_K} \footnote{Note that the number of 2.8 million kaons quoted in Ref.~\cite{comp_K} is incorrect; the value after all selections was about 0.6 million.}.

\subsection{Particle identification}

Hadron identification was accomplished using the RICH \cite{comp_exp} detector, as detailed in Ref.~\cite{comp_pi}.
A summary of the most important aspects is given here.
The method employed relies on a maximum likelihood approach using the pattern of photons detected by the RICH detector. Likelihood values are computed by comparing the observed photo-electron pattern with the expected patterns corresponding to different mass hypotheses ($m_{\pi}$, $m_{\rm K}$, $m_{\rm p}$, $m_{\rm e}$), also accounting for the distribution of background photons.

The yields of identified hadrons denoted by $N_{\text{true}}^i$ are derived by applying an unfolding algorithm to the yields $N_{\text{rec}}^j$ of reconstructed hadrons:
\begin{equation}
    N_{\text{true}}^i = \sum_j (P^{-1})_{ij} \cdot N_{\text{rec}}^j
\end{equation}
with $i,j = \{\pi, \rm{K}, \rm{p}\}$.
Here, the so-called purity--efficiency matrix $P_{ij}$ serves to correct hadron yields by accounting for particle identification and misidentification.
Its diagonal elements correspond to the efficiency of identification, while its
off-diagonal elements represent probabilities of misidentification.
The elements of this matrix are determined using data samples containing pions, kaons and protons originating from two-charged-particle decays of K$^{0}_{\rm {s}}$, $\phi$ and $\Lambda$ respectively.
All these matrix elements depend primarily on the values of $p$ and $\theta_{\rm{RICH}}$ at the entrance point of the particle into the RICH detector.
In total, about 30 $P_{ij}$ matrices are extracted from data, forming a two-dimensional grid of $p$
and $\theta_{\rm{RICH}}$.
The elements of $P_{ij}$ are found to be consistent for
the two electric charges of
the studied hadrons and are averaged in the analysis.

\begin{figure}
    \centerline{
    \includegraphics[clip,width=0.5\textwidth]{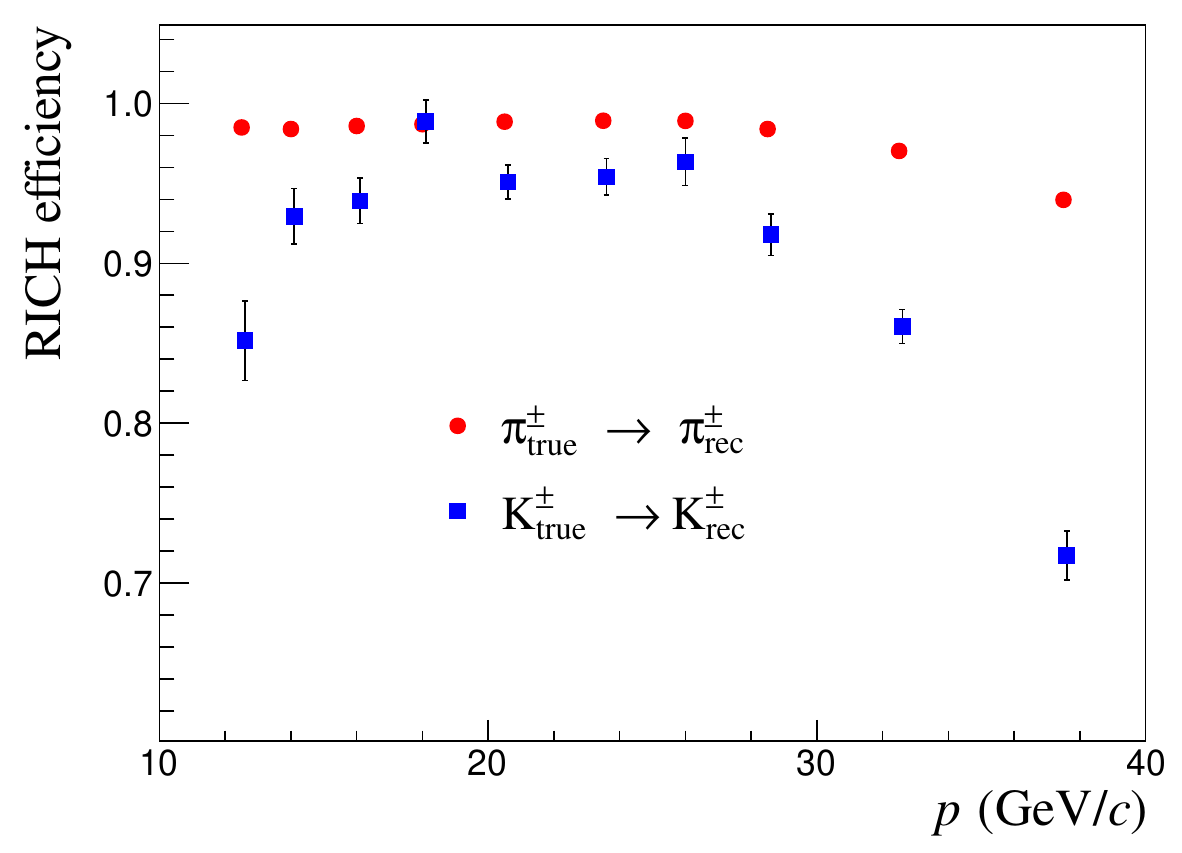}
    \includegraphics[clip,width=0.5\textwidth]{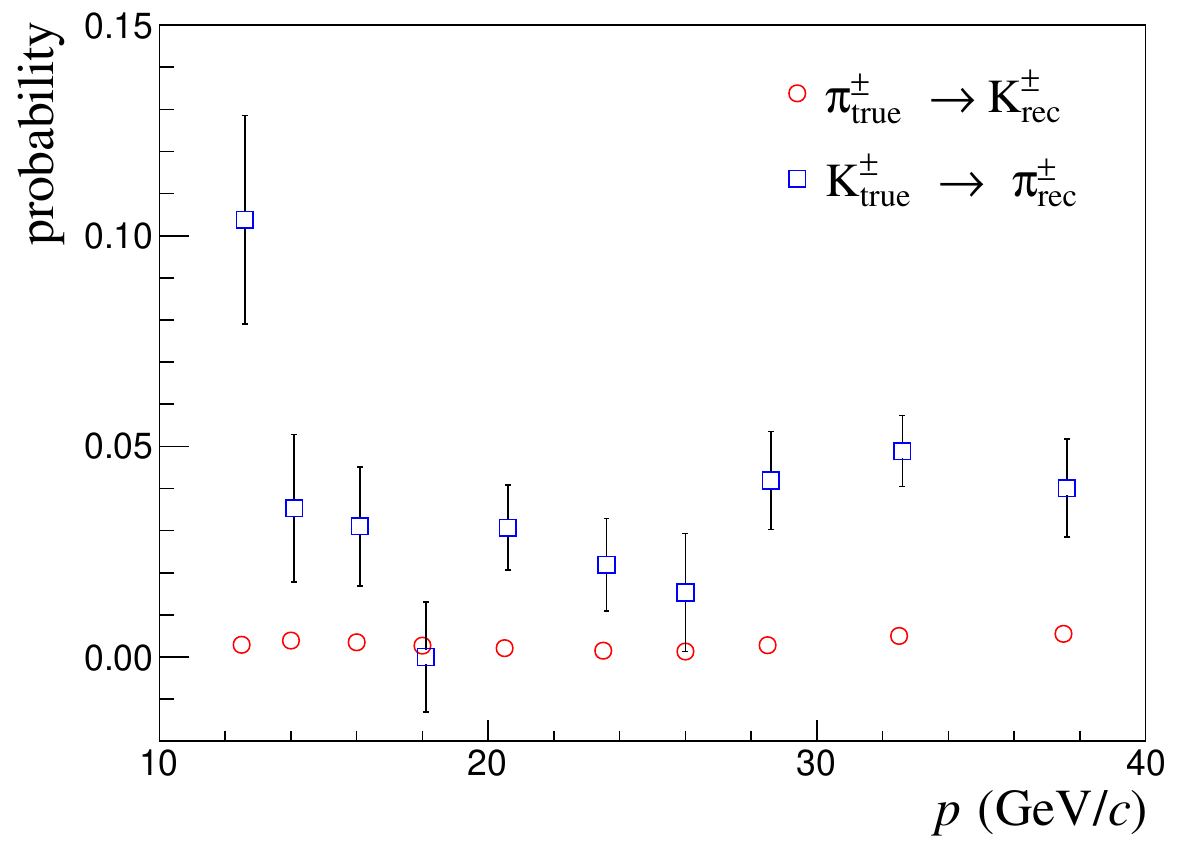}}
    \caption{Momentum dependence of the purity--efficiency matrix elements for $0.010~<~\theta_{\rm{RICH}}~<~0.032$. \\ Left panel: efficiency of $\pi^{\pm}$,
    K$^{\pm}$ identification, right panel: corresponding misidentification probabilities.}
    \label{fig:rich}
\end{figure}

The efficiency for pion identification ranges from 94\% to 99\% with a median value of 98\%. Misidentification of pions as kaons is rare, remaining below 0.6\% with a median of 0.3\%.
The efficiency for kaon identification varies from 71\% to 99\% with a median of 95\%. The misidentification of kaons as pions is below 15\% with a median of 2.4\%.
Examples of the momentum dependence for the purity--efficiency matrix elements are presented in Fig.~\ref{fig:rich} for $0.010<\theta_{\rm{RICH}}<0.032$.

\subsection{Acceptance correction}

The corrections for geometric and kinematic acceptance,
including detector inefficiencies, resolutions and bin migration within the COMPASS apparatus, are determined using a Monte Carlo (MC) simulation.
This simulation successfully reproduces within 20\% the kinematic distributions observed in the experimental data by use the following tools:
 $i)$~LEPTO \cite{lepto} for the generation of DIS events,
   $ii)$~JETSET \cite{lund} for hadronisation (tuned as in Ref.~\cite{comp_dg}),
   $iii)$~GEANT4 \cite{geant4} to describe the spectrometer and for modelling secondary hadron interactions.

The acceptance correction is calculated within narrow regions of $(x, y, z)$,
thereby mitigating its sensitivity to the specific choice of the
used generator. It is given by the ratio of reconstructed to generated multiplicities:

\begin{equation}
A(x, y, z) = \frac{{\rm d}N^i_{\rm rec}(x, y, z)/N^{\rm DIS}_{\rm rec} (x, y)}
{{\rm d}N^i_{\rm gen}(x, y, z)/N^{\rm DIS}_{\rm gen}(x, y)} .
\end{equation}

In this equation, the values of generated kinematic variables are employed for the generated events,
while the values of reconstructed kinematic variables are used for the reconstructed events.
The reconstructed hadrons in MC are subject to the same kinematic and geometric
selection criteria as the actual data, whereas the generated hadrons are subject solely to
kinematic requirements. Events generated outside the acceptance region and reconstructed inside
are included in the calculations.
The average value of the acceptance is approximately  70\% to 80\%. It is nearly constant across different regions of $x$, $y$ and $z$,
except for the high-$x$ and low-$y$ region, but it always
exceeds 25\%.

\begin{figure}
    \centering
    \includegraphics[clip,width=0.8\textwidth]{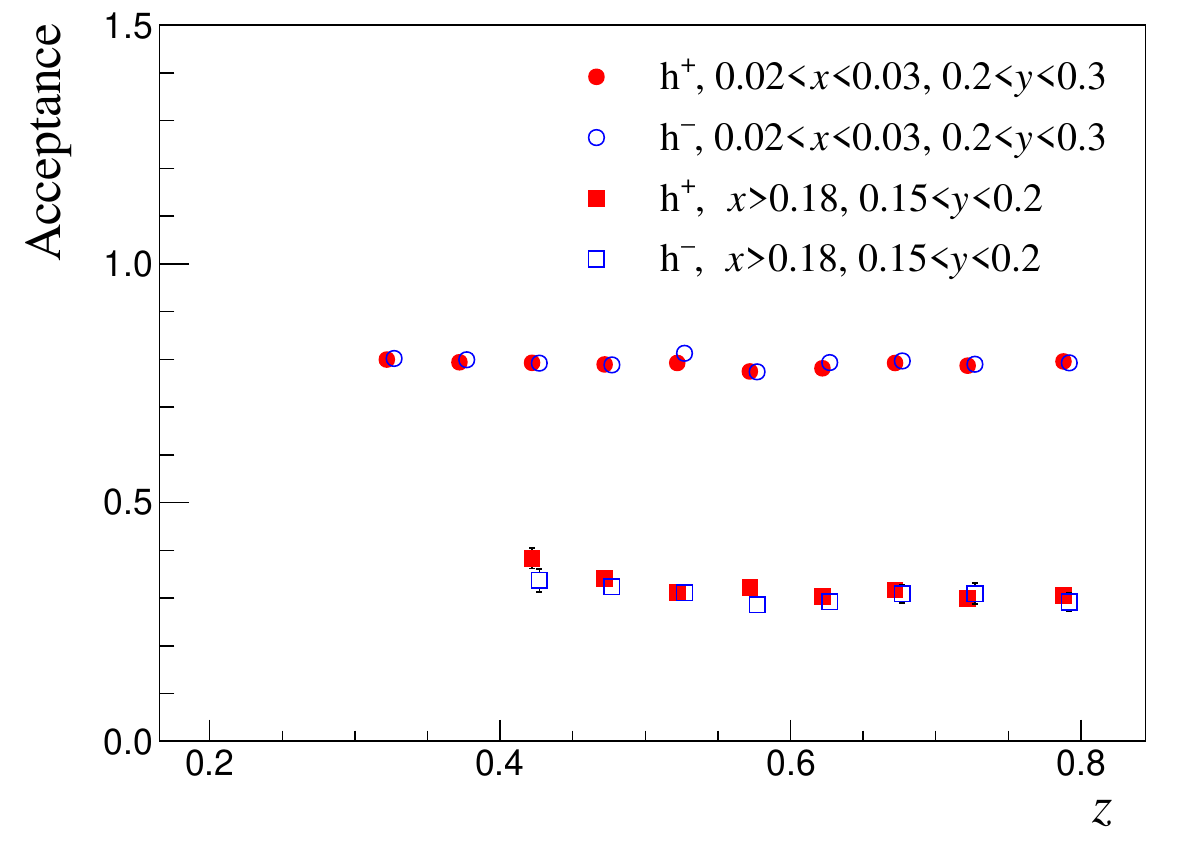}
    \caption{Example of COMPASS acceptance in two selected $(x,y)$ bins as a function of $z$ (for clarity staggered horizontally)}.
    \label{fig:acc}
\end{figure}

\subsection{Vector-meson correction}

A fraction of the mesons measured in SIDIS originates
from diffractive production of vector-mesons that subsequently decay into lighter
hadrons. This fraction represents a higher-twist contribution
to the SIDIS cross section \cite{hermes}. It cannot be adequately described by the QCD parton model
with the independent fragmentation mechanism encapsulated in FFs.
Hence, extracting FFs
from data that includes this fraction would introduce a bias,
and undermine the universality of the model.

In order to correct the multiplicities for the vector-meson (VM) contribution, we estimate the fraction of final-state hadrons
originating from the diffractive decay of $\rho^0$ and $\phi$ mesons,
while other vector-mesons characterised by
smaller production cross sections and multi-body decay channels are not considered.
In a given $(x, y, z)$ bin, two fractions are obtained: one for the hadron sample and one for the DIS sample. Their ratio,
VM$_{\rm corr}^{i}(x,y,z)$/VM$_{\rm corr}^{\rm DIS}(x,y)$,
where $i=\{ \pi^{+}, \pi^{-}$, K$^{+}$, K$^{-}$, h$^{+}$, h$^{-}$ \}  represents the VM contribution used to correct the multiplicity values given in Eq~(\ref{eq:mult_data}).
The estimation of the correction relies on two Monte Carlo simulations.
To simulate SIDIS events the LEPTO~6.5 MC
generator is used with switched-off diffractive contributions.
Additionally, the HEPGEN \cite{hepgen} generator is used to simulate the diffractive production of
$\rho^0$ and $\phi$ mesons.
Unlike in previous COMPASS multiplicity analyses, the spectra of missing energy obtained from HEPGEN and LEPTO are
normalised to the data to ensure a better description of data by Monte Carlo. This approach reduces our dependence on the Goloskokov--Kroll model \cite{hepgen_model}
for the cross section used in HEPGEN.

\begin{figure}
    \centerline{
    \includegraphics[clip,width=0.5\textwidth]{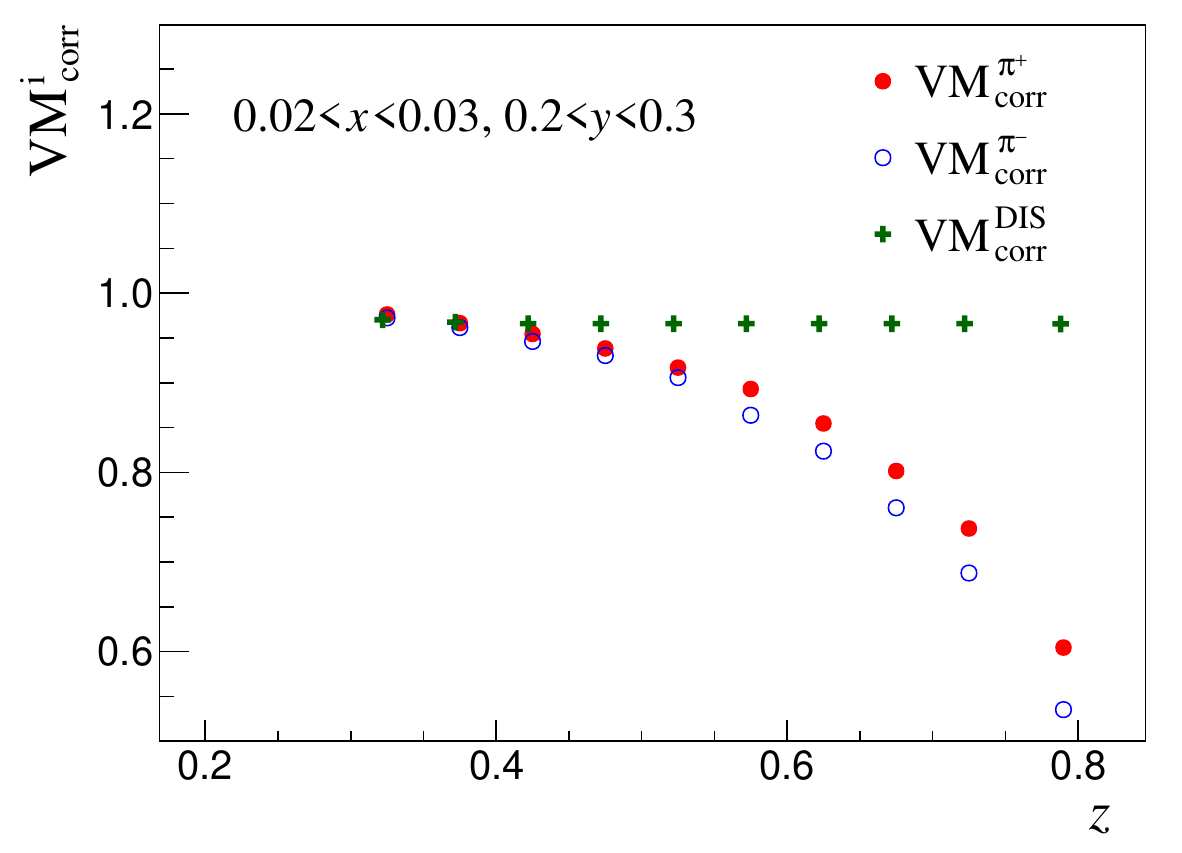}
    \includegraphics[clip,width=0.5\textwidth]{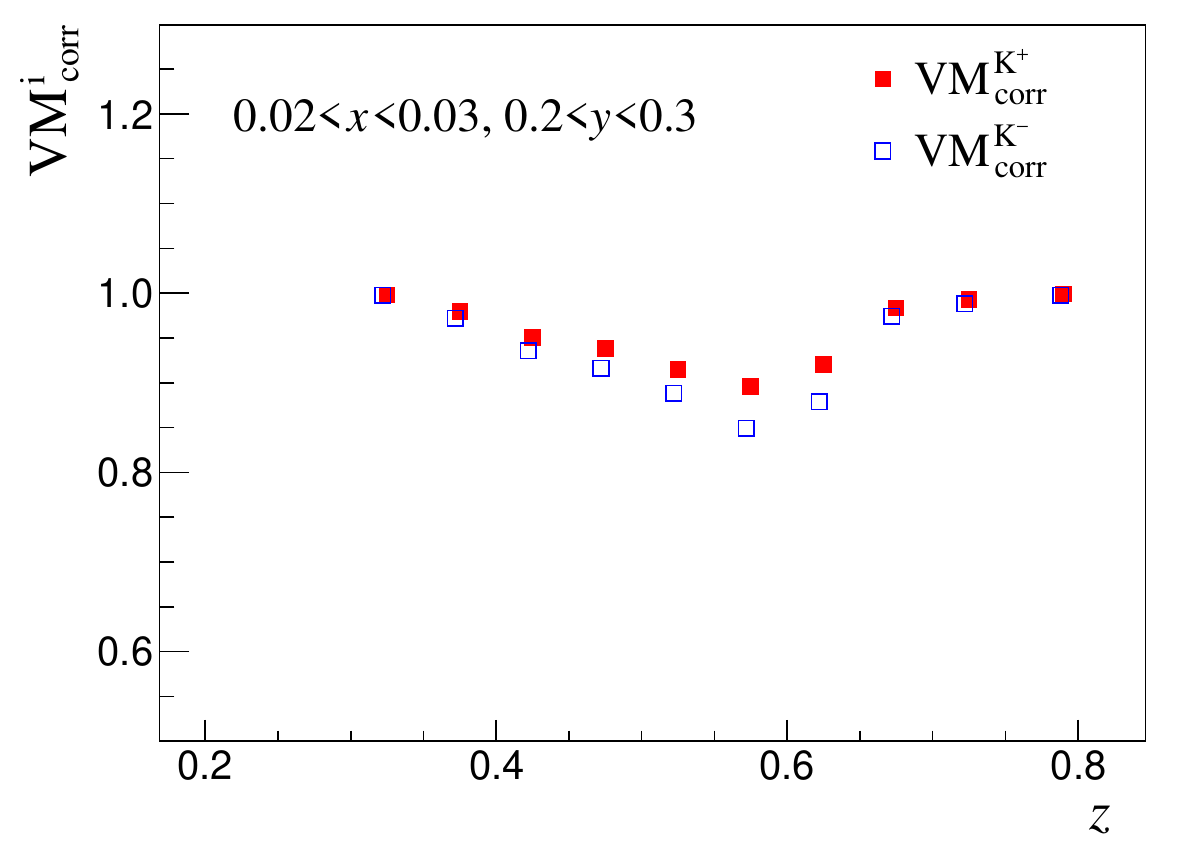}}
    \caption{Values of VM$_{\rm corr}$ in a selected bin of $(x,y)$ as a function of $z$ (for clarity staggered horizontally) for pions and DIS events (left), and kaons (right).}
    \label{fig:vm}
\end{figure}

Diffractive events are particularly prominent at low values of $x$ and $Q^2$.
Concerning pions and unidentified hadrons, the primary contribution arises from the decay of $\rho^0$ into two charged pions.
This contribution is most significant at high values of $z$, where our simulations indicate that it exceeds the SIDIS cross section itself.
Concerning kaons, the primary contribution originates from
the decay of $\phi$ into two charged kaons.
Given the relatively low $\phi$-mass in comparison
to the threshold mass of two kaons, the resulting decay products have very low momenta in the centre-of-mass of the $\phi$.
Consequently, within the COMPASS kinematics, these events are predominately observed in the $0.4 < z < 0.6$ range.
The maximum correction is about 24\% for $z \approx 0.6$
and $Q^2 \approx 1$ (GeV/$c)^2$.
The obtained corrections are shown in Fig.~\ref{fig:vm} in a selected bin of $(x,y)$
as a function of $z$, for different particle types and electric charges
as well as for inclusive DIS events.
These corrections are comparable to those used in our earlier analysis of data taken with an isoscalar target.

\subsection{Radiative corrections}
In previous COMPASS multiplicity analyses,
corrections for higher-order QED effects were computed using the program TERAD, based on the scheme described in Ref.~\cite{terad}.
This approach is strictly valid only for inclusive DIS events.
 To estimate corrections to the yields of $\pi$, K, h, the TERAD calculations were performed excluding the elastic and quasi-elastic radiative tails.
 This was an approximation, as TERAD could not account for a $z$ dependence in the cross section neither for different particle types and electric charges.

\begin{figure}
\centerline{\includegraphics[clip,width=0.5\textwidth]{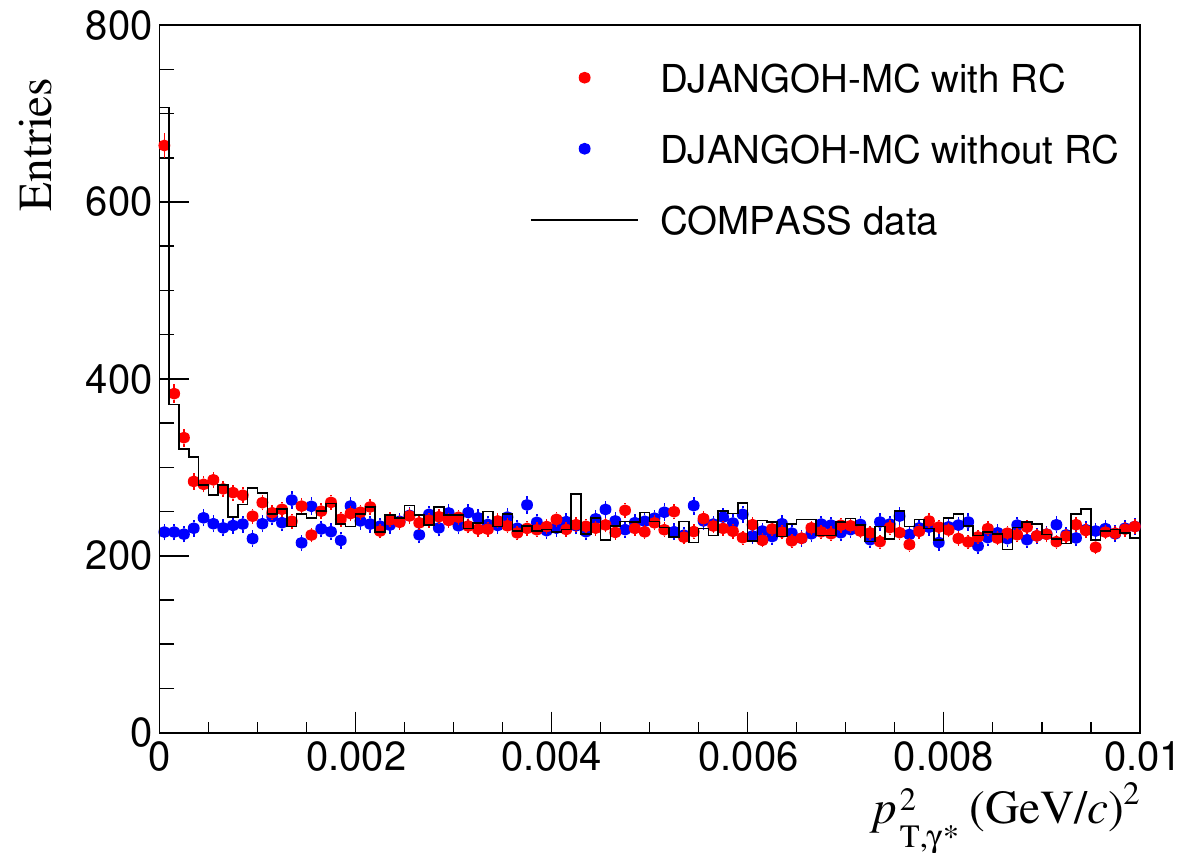}
\includegraphics[clip,width=0.5\textwidth]{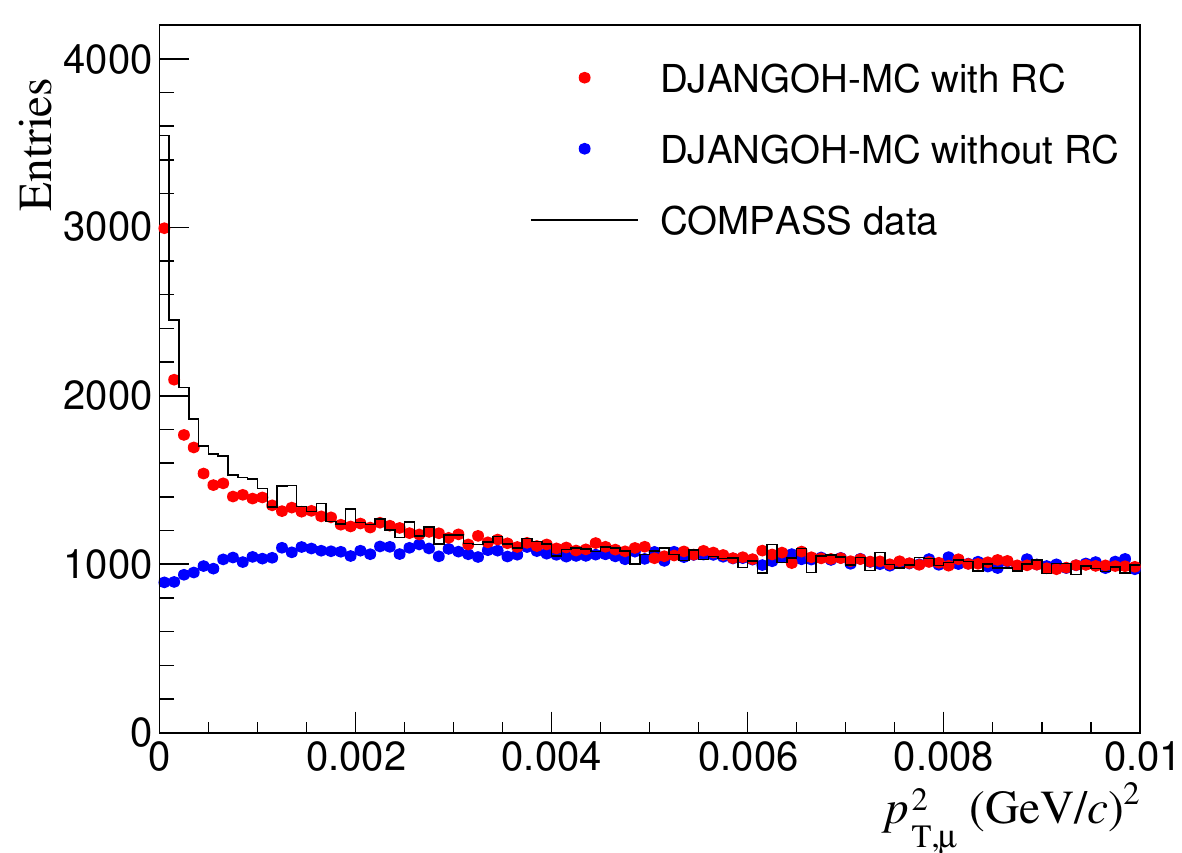}}
\centerline{\includegraphics[clip,width=0.5\textwidth]{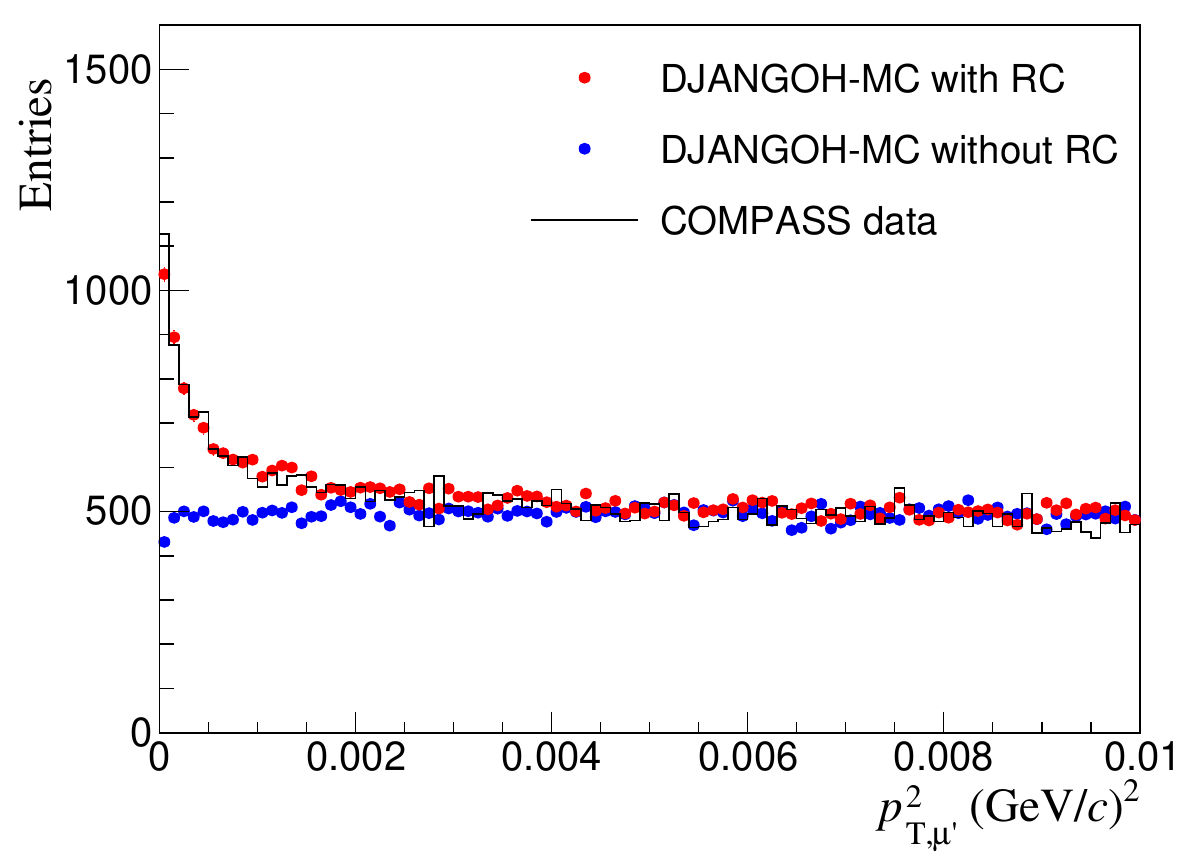}
\includegraphics[clip,width=0.5\textwidth]{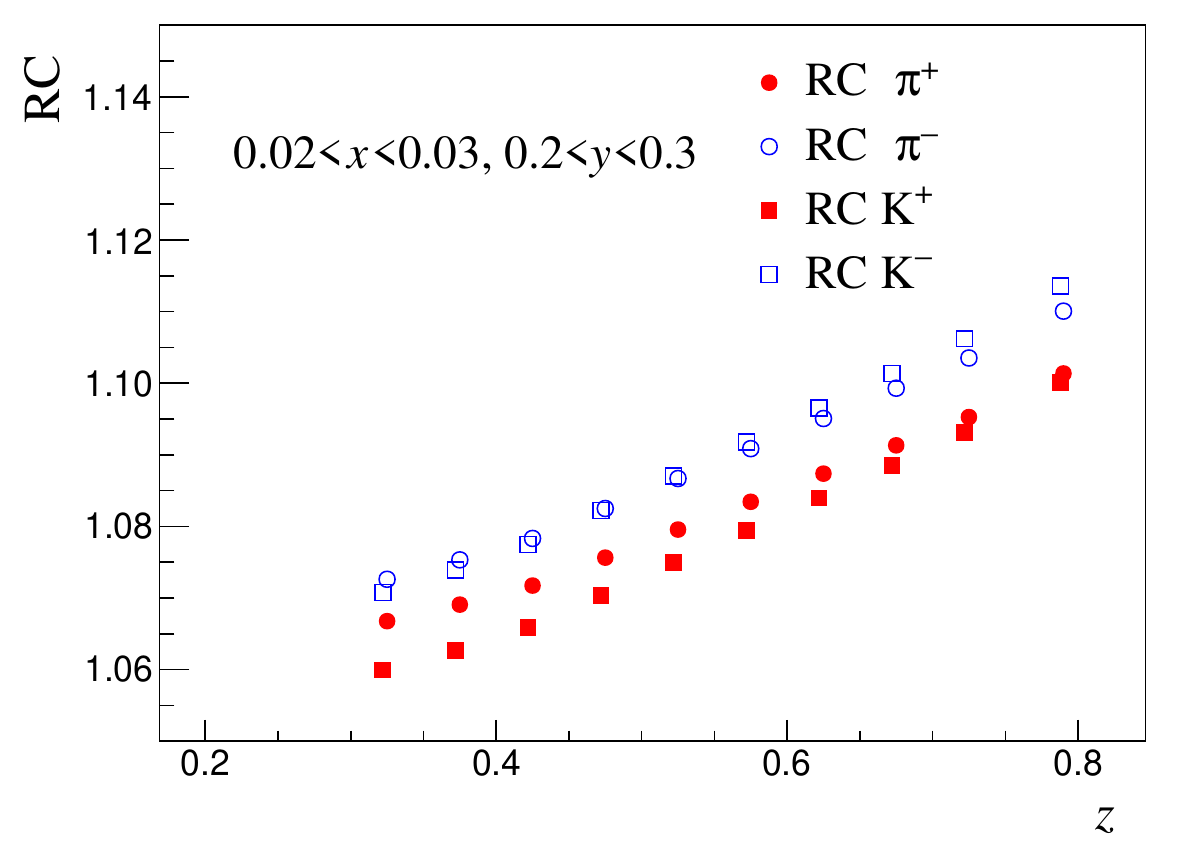}}
\caption{Transverse momentum of hadron candidates with respect to the direction of virtual-photon (top-left), and of incoming (top-right) and outgoing (bottom-left)  muon. A clear peak from electrons (positrons) coming from $\gamma$ conversion is seen. Experimental data are compared to DJANGOH-MC results with and without radiative effects. In the bottom-right panel the RC values for $\pi^{\pm}$ and ${\rm K}^{\pm}$ are shown in a selected $(x,y)$ bin as a function of $z$ (for clarity staggered horizontally)}. \label{fig:rc1}

\end{figure}

For the present analysis, COMPASS uses the event generator DJANGOH  \cite{djangoh}, referred to as DJANGOH-MC in the following.
The DJANGOH-MC results agree well with those from TERAD for inclusive DIS. They also reproduce reasonably well the measured data.
In Fig.~\ref{fig:rc1},
examples of data distributions
are shown in comparison to DJANGOH-MC results with and without radiative effects.
While most of the radiation is produced along the directions of the incoming and outgoing muon, the non-negligible mass of the muon
induces as the third emission direction that of the virtual-photon.
The top-left panel of Fig.~\ref{fig:rc1}  shows distributions of the squared transverse momentum of hadron candidates with respect to the real-photon direction $p_{\rm T,\gamma^*}^2$.
A clear peak is visible for very low values of $p_{\rm T,\gamma^*}^2$ both in the data and in the DJANGOH-MC distributions. In fact, the observed peak originates from real-photon conversion into e$^+$e$^-$ pairs and subsequent misidentification of electron or positron as charged hadron candidate.
In the top-right and bottom-left panels of Fig.~\ref{fig:rc1}, the same comparison is presented as
in the top-left panel, but the transverse momentum is measured with respect to the incoming muon (top-right panel)
or outgoing muon (bottom-left panel).
In all three cases, a reasonable agreement between data and DJANGOH-MC is observed.

In order to account for radiative effects in the multiplicity results, a correction factor denoted as RC is calculated in bins of $(x,y,z)$, using a two-step procedure. In the first step, the multiplicities are obtained using DJANGOH-MC with and without radiative effects, yielding the RC value as their ratio.
However, for inclusive DIS the TERAD calculation yields a more complete description of radiative effects.
Hence the RC value obtained in the first step above is multiplied
in the second step by the TERAD vs. DJANGOH-MC ratio of radiative corrections for inclusive DIS events.
The correction applied in the second step typically remains below 2\% within the kinematic range of the present analysis.
An illustration of the derived values of RC within a chosen $(x, y)$ bin is shown as a function of $z$ in the bottom-right panel
of Fig.~\ref{fig:rc1}
for $\pi^{\pm}$ and K$^{\pm}$. A distinct $z$ dependence is evident with a slight variation of the RC values observed for different particle types and electric charges.

\begin{figure}
\centerline{\includegraphics[clip,width=0.8\textwidth]{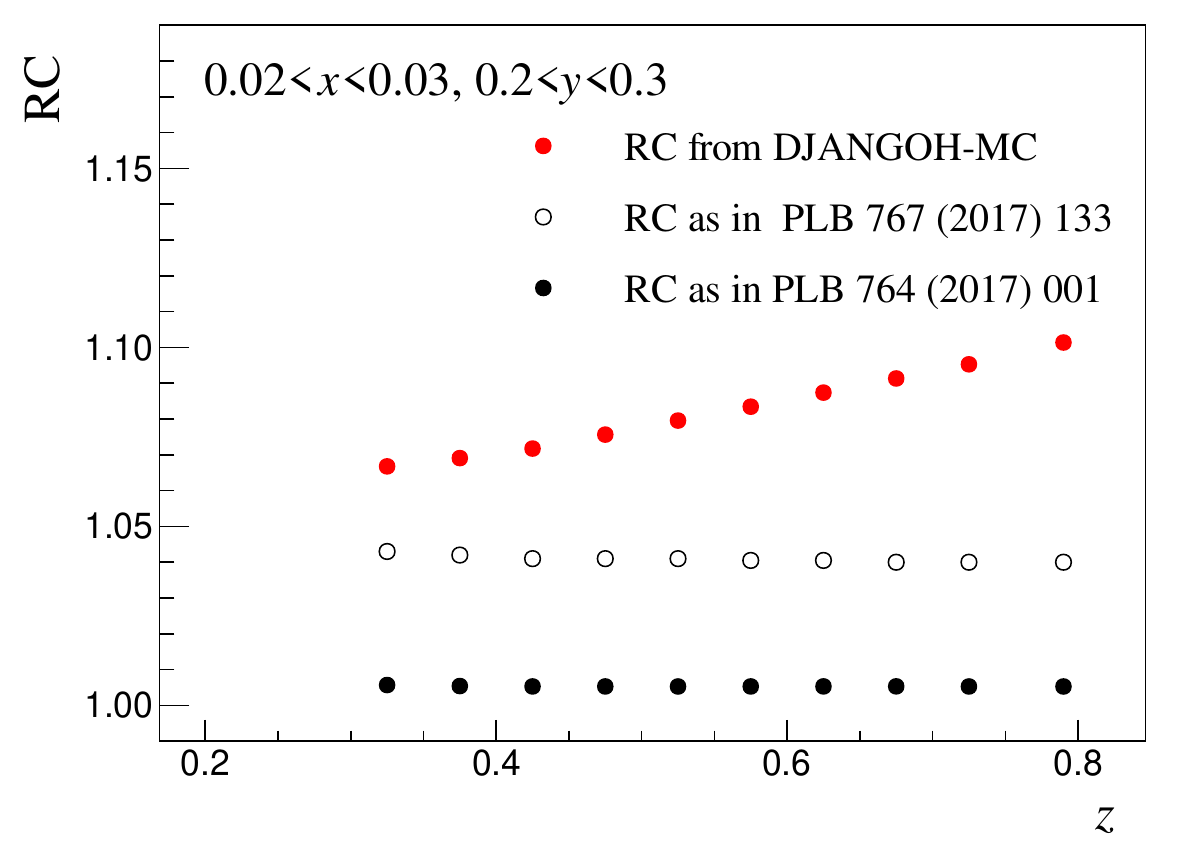}}
\caption{Comparison of radiative corrections for positive pions for $0.02<x<0.03$ and $0.2<y<0.3$ as a function of $z$ following various procedures
$i)$ using DJANGOH-MC, $ii)$ as in Ref.\cite{comp_K}, $iii)$ as in Ref. \cite{comp_pi}.    }
\label{fig:rc2}
\end{figure}

Figure~\ref{fig:rc2} shows a comparison of the present RC values for pions
with the ones used in the previous multiplicity analyses of Refs~\cite{comp_pi} and~\cite{comp_K}
for the same bin as used in Fig.~\ref{fig:rc1}. The actual correction is seen to be 3\% to 6\% larger than that used in Ref.~\cite{comp_K}, which itself was already about 4\% larger than that used in Ref.~\cite{comp_pi}, with the two older corrections having almost no $z$-dependence. Within the kinematic domain of the COMPASS multiplicity analysis, the earlier corrections appear underestimated by up to 12\%. The main reason for this change is a reduction of the radiative corrections for events with hadrons, if the hadron phase space is properly taken into account. In Ref.~\cite{comp_K} an approximation of this effect was applied, leading to larger corrections. Updated radiative corrections for the results obtained with the isoscalar target \cite{comp_pi,comp_K}  will be provided in a separate paper.

\subsection{Systematic studies} \label{sec:an:sys}

The primary sources of systematic uncertainties in the multiplicity analysis are associated to the
evaluation of acceptance, RICH performance, diffractive vector-meson contributions and QED radiative effects.

Similar studies as described in Ref.~\cite{comp_pi} are done to estimate the acceptance uncertainty. Additional
studies, $e.g.$ utilising DJANGOH-MC
simulations with and without radiative effects are also performed. The resulting relative uncertainty is 4\% as in the previous multiplicity analyses.

In order to estimate the uncertainty related to the RICH identification and the unfolding procedure, different RICH purity--efficiency matrices are
constructed by varying the values of matrix elements within
their statistical uncertainties. However, the statistical limitations in the 2016 data,
including lower integrated luminosity and the absence of a specific low $Q^2$ trigger,
lead to statistical uncertainties in the extracted parameters higher than in Refs \cite{comp_pi, comp_K}.
Additionally, variations in the number of
photons in the RICH are observed, which depend on the radial distance
of the track from the beam axis at the RICH entrance.
One reason is the length and position of the target,
which was designed for deeply virtual Compton scattering measurements.
Altogether, the relative systematic uncertainty
associated with the RICH is higher than that for the 2006 data analysis and is estimated to be between 3\% and 10\%.

In the case of the vector-meson cross section uncertainty,
the normalisation of the LEPTO and HEPGEN MC simulations to real data distributions provides
increased confidence in the obtained results.
Consequently, the relative uncertainties of vector-meson cross sections are reduced to
about 20\%. This leads to a relative uncertainty of pion and unidentified hadron multiplicities,
which in a few bins at low $x$, low $Q^2$ and  high $z$
can reach up to 23\%, while it is up to 5\% for kaon multiplicities.
All vector-meson corrections and their uncertainties strongly depend on the kinematic region.

For the radiative correction uncertainty,
the studies comparing data and Monte Carlo simulations suggest an uncertainty of about
$0.25|RC-1|$. In addition, a constant term of 2\% is added in quadrature to
account for differences observed between DJANGOH-MC and TERAD for the inclusive DIS correction.

Combining in quadrature all individual contributions to the systematic uncertainty yields the total systematic uncertainty $\sigma_{\rm syst}$, which varies between 5\% and 24\%
depending on the kinematic region, with a median value around 6\%.
This systematic uncertainty is generally larger than the statistical one
at low values of $z$ and low values of $x$, while at higher values of $z$ or $x$
the statistical uncertainty dominates.
It is worth noting that a large fraction (about 80\%) of the total systematic uncertainty is estimated to be
bin-to-bin correlated, and the remaining fraction ($\sqrt{1 - 0.8^2}\sigma_{\rm syst}$) is
uncorrelated.

\section{Results}

The multiplicities of charged hadrons (h$^{\pm}$), pions ($\pi^{\pm}$) and kaons (K$^{\pm}$)
presented in the following figures were corrected for diffractive contributions
and QED radiation effects unless otherwise specified.
The numerical values of these multiplicities can be found on HEPData \cite{hep_data},
along with the corresponding correction values. The same binning
as in previous analyses \cite{comp_pi, comp_K} is used for $x$, $y$ and $z$, see Table~\ref{tab:res_1}.
The $Q^2$ values span from 1 (GeV/$c)^2$ at the lowest $x$ to approximately 60 (GeV/$c)^2$
at the highest $x$ with an average value of about 3~(GeV/$c)^2$.
In total, this analysis yields about 300 data points for each hadron species and electric charge.

\begin{table}[ht]
    \centering
    \caption{Lower and upper bin limits for the $(x, y, z)$ binning}
    \begin{tabular}{c c c c c c c c c c c c c c}
       \hline
         &   bin limits \\  \hline
        $x$ &  0.004 & 0.01 & 0.02 & 0.03 & 0.04 & 0.06 & 0.10 &  0.14 & 0.18 & 0.40            \\
        $y$ &  0.10 & 0.15 & 0.20&  0.30 & 0.50 & 0.70            \\
        $z$ &  0.20 & 0.25 & 0.30 & 0.35 & 0.40 & 0.45& 0.50 & 0.55 & 0.60 & 0.65 & 0.70 & 0.75 & 0.85 \\   \hline
    \end{tabular}
       \label{tab:res_1}
\end{table}

\begin{figure}
    \centering
    \includegraphics[clip,width=1.0\textwidth]{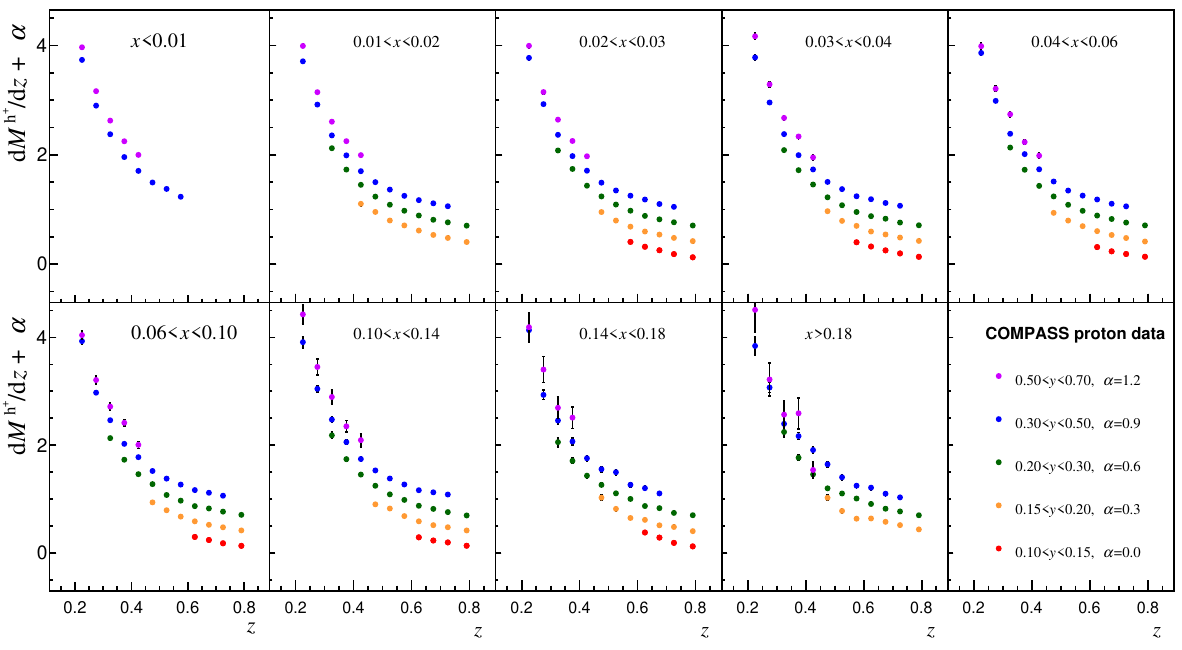}
    \caption{Positive hadron multiplicities versus $z$ for nine $x$ bins and five $y$ bins (for clarity staggered vertically by $\alpha$). Only statistical uncertainties are shown.}
    \label{fig:res-1}
\end{figure}

\begin{figure}
    \centering
    \includegraphics[clip,width=1.0\textwidth]{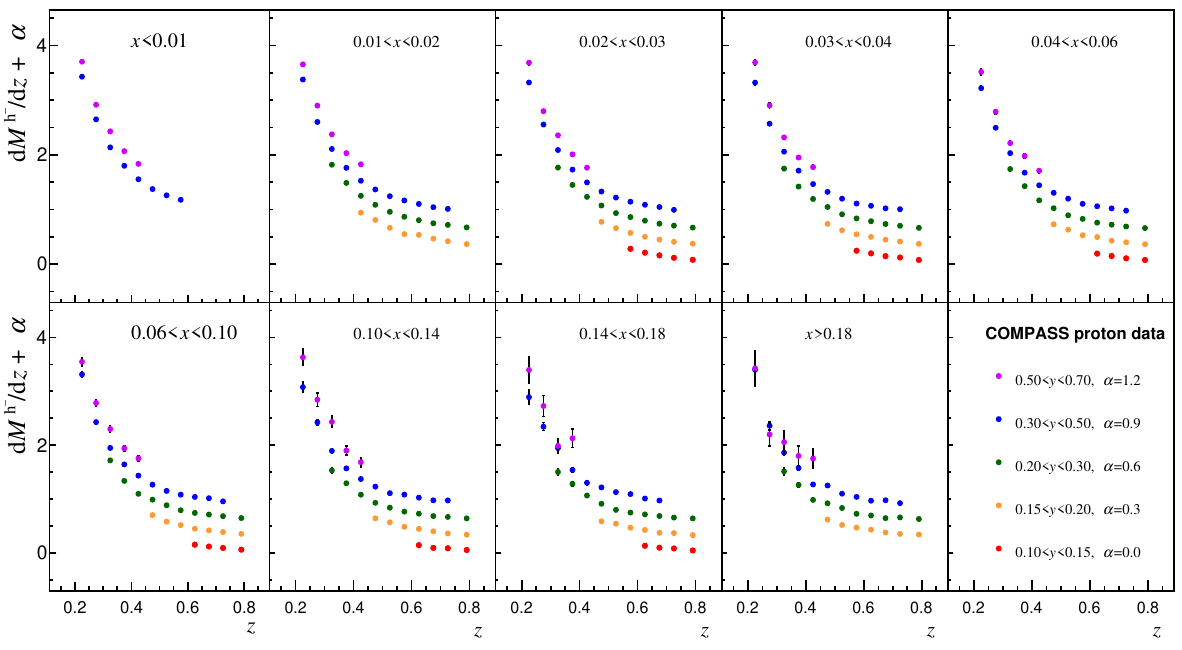}
    \caption{Same as Fig. \ref{fig:res-1} for negative hadrons.}
    \label{fig:res-2}
\end{figure}

The results for unidentified hadrons and pions, which exhibit similar
characteristics, are discussed together, while the discussion of kaon results follows below. In Figs~\ref{fig:res-1} and~\ref{fig:res-2} multiplicity results for
unidentified hadrons of both electric charges are presented.
These figures illustrate the multiplicity values as a function of $z$ in nine $x$-bins.
In each panel, the results from five $y$-bins are displayed in a staggered manner. The uncertainties in these figures are statistical ones, which are often smaller than the size of the data points.
As expected, the multiplicities of h$^+$ are larger than those of h$^-$, mainly due to
u-quark dominance in the interaction. This difference becomes more pronounced for higher $x$ values.
With increasing $z$, the multiplicities decrease, and the distinction between h$^+$ and h$^-$ becomes more evident. This phenomenon is related to the fact that
$D_{\rm unf}$ is expected to decrease more rapidly with growing $z$ than $D_{\rm fav}$, see $e.g.$ Ref. \cite{dss_01}.

\begin{figure}
    \centering
    \includegraphics[clip,width=1.0\textwidth]{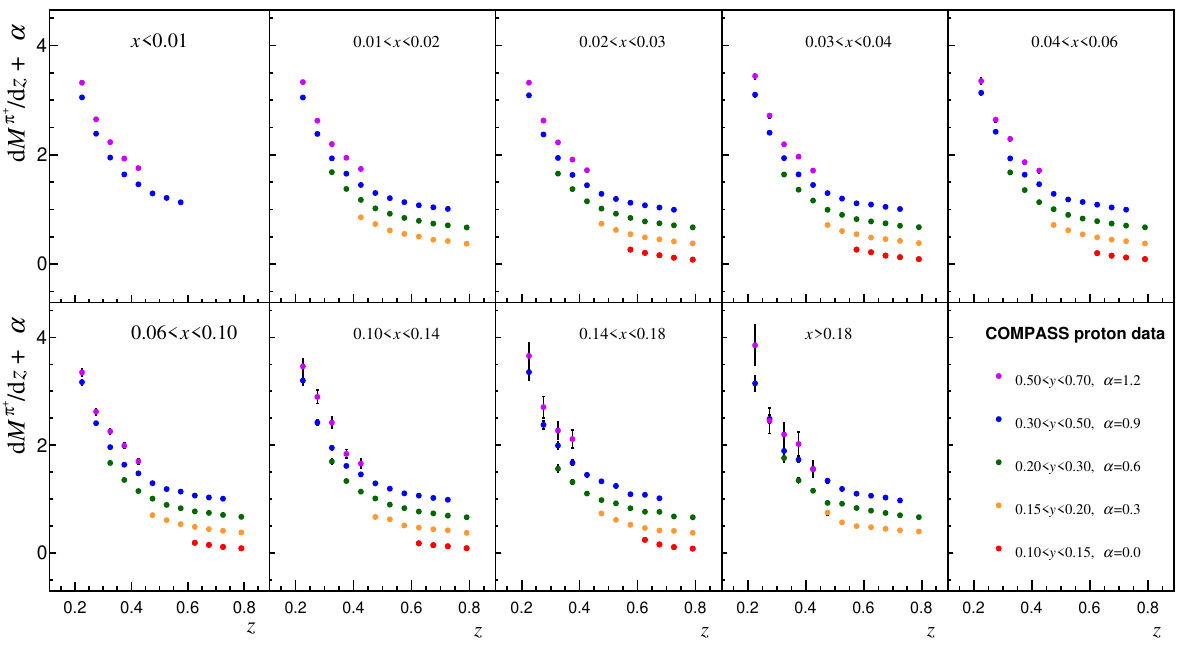}
    \caption{Same as Fig. \ref{fig:res-1} for positive pions.}
    \label{fig:res-3a}
\end{figure}

\begin{figure}
    \centering
    \includegraphics[clip,width=1.0\textwidth]{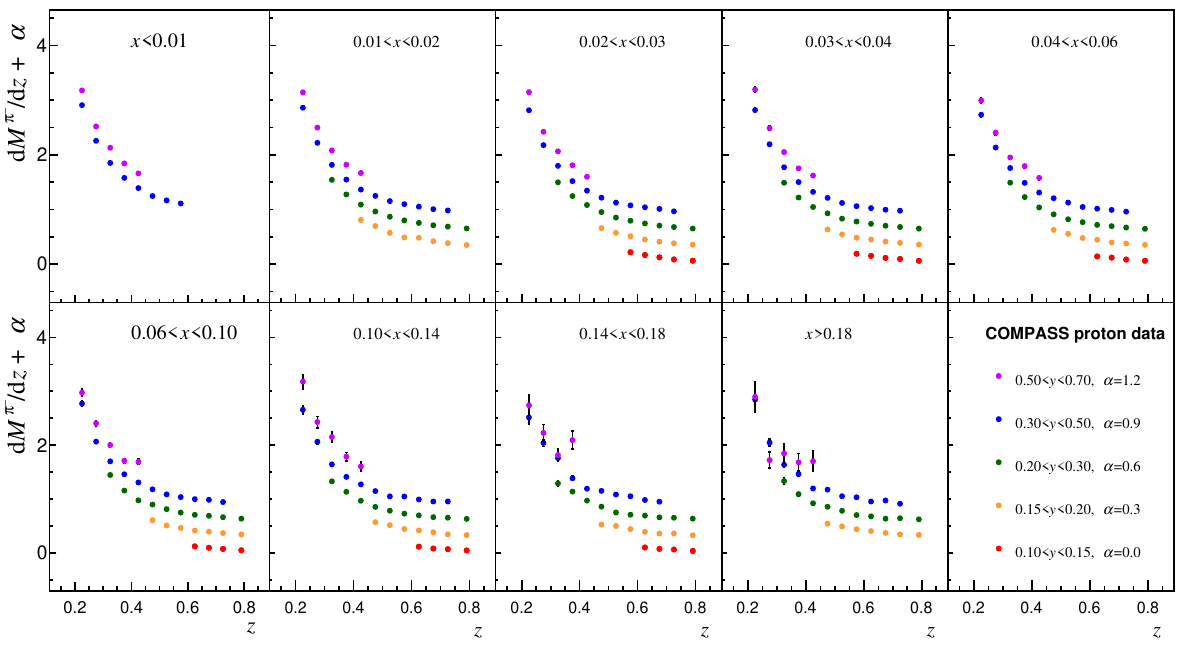}
    \caption{Same as Fig. \ref{fig:res-1} for negative pions.}
    \label{fig:res-3b}
\end{figure}

\begin{figure}
    \centering
    \includegraphics[clip,width=1.0\textwidth]{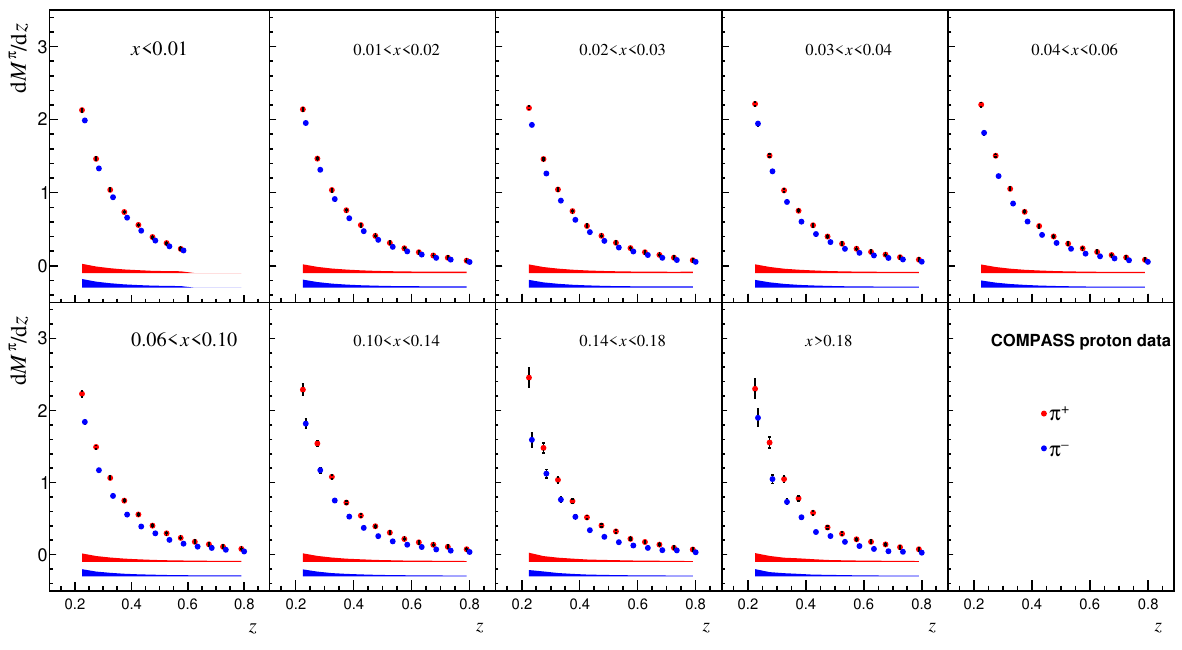}
    \caption{Positive (red) and negative (blue) pion multiplicities versus $z$ (for clarity staggered horizontally) for nine $x$ bins.
 The data points are shown with statistical uncertainties,
 while the bands indicate systematic uncertainties.}
    \label{fig:res-3}
\end{figure}

The results for pion multiplicities are presented in Figs~\ref{fig:res-3a} and~\ref{fig:res-3b}
in the same staggered way as for the unidentified hadrons discussed above.
In addition, in Fig.~\ref{fig:res-3} the pion multiplicities are shown
as a function of $z$ in nine $x$-bins averaged over $y$.
The data points in Fig.\ref{fig:res-3}  are shown together with statistical uncertainties,
and the shaded bands represent the total systematic ones; the data for negative electric charge are staggered by 0.01 in $z$.
The distinction between positive and negative pions is less pronounced compared to the
results for hadrons.

The results presented in Fig.~\ref{fig:res-3} can be further integrated over $z$ for both electric
charges. This integration yields the quantity $\mathcal{M}^{\pi^+}\!\!+\!\mathcal{M}^{\pi^-}$.
This sum of multiplicities, integrated over $z$ and averaged over $y$, has several interesting
features.
For instance, in LO pQCD following Eqs.(\ref{eq:1})-(\ref{eq:3})
one obtains that: $\mathcal{M}^{\pi^+}\!\!+\!\mathcal{M}^{\pi^-}$ =
$\mathcal{D}_{\rm fav}\!+\!\mathcal{D}_{\rm unf} + {s_{\rm corr}}$.
Here $\mathcal{D}$ represents the $z$-integrated value of $D$, $s_{\rm corr}$ is associated with
the strange quark contribution and anticipated to be small.
As FFs do not depend on $x$, we do not expect any $\mathcal{M}^{\pi^+}\!\!+\!\mathcal{M}^{\pi^-}$
dependence on $x$. However, there is a correlation between $x$ and
$Q^2$ in fixed-target kinematics. Since FFs depend on $Q^2$, this $(x,Q^2)$ correlation may indirectly
introduce a weak $x$-dependence.
Furthermore, the same
leading-order pQCD calculations performed for the isoscalar target yield nearly identical results,
namely $\mathcal{M}^{\pi^+}\!\!+\!\mathcal{M}^{\pi^-}\!\approx \mathcal{D}_{\rm fav}\!+\!\mathcal{D}_{\rm unf}$ + s'$_{\rm corr}$, see \cite{comp_pi}. Thus, in LO pQCD
this sum of multiplicities is expected to be nearly the same for the proton and isoscalar target measurements as the difference between s$_{\rm corr}$ and
s'$_{\rm corr}$ is anticipated to be below 1\%.

\begin{figure}
    \centerline{
    \includegraphics[clip,width=0.5\textwidth]{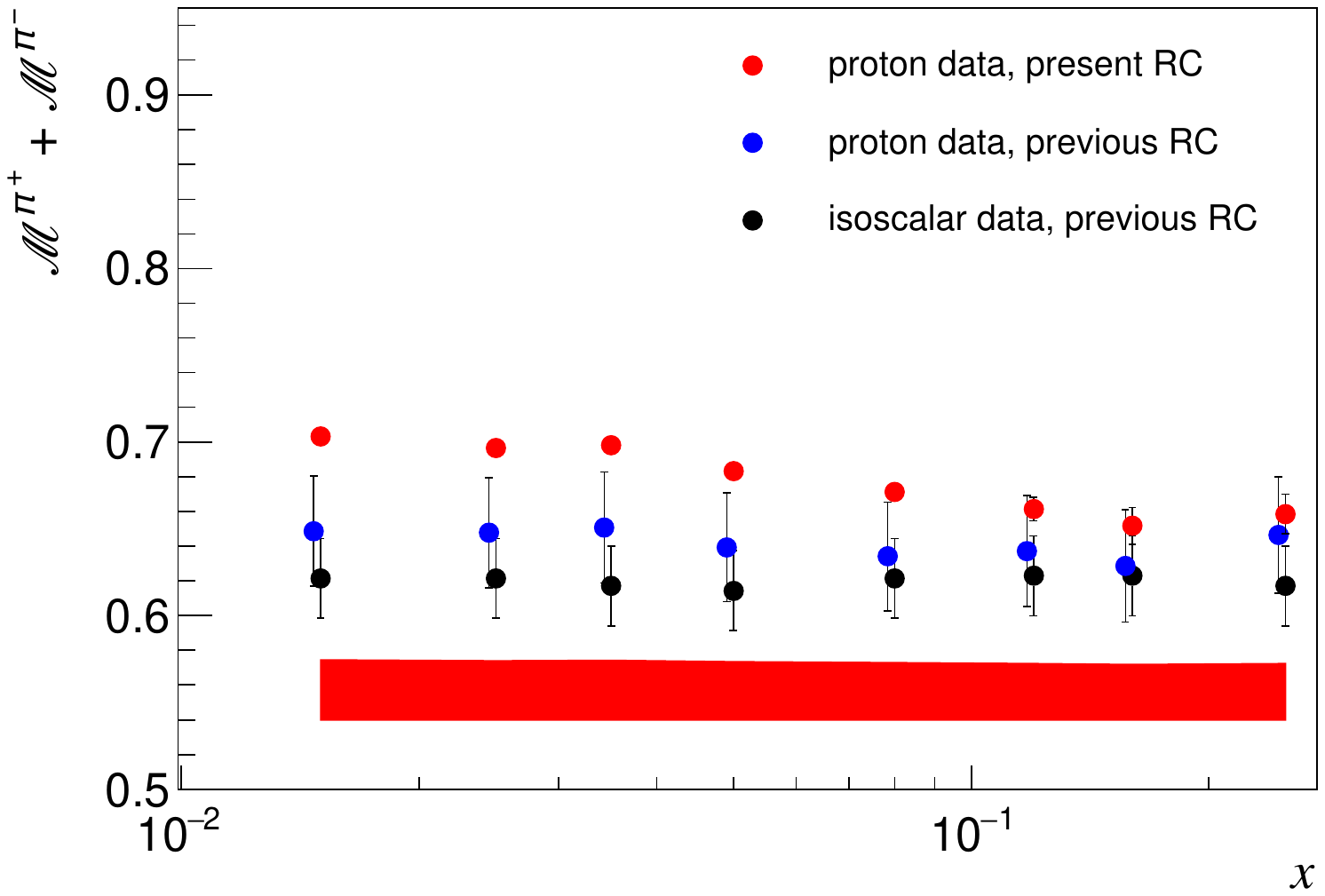}
    \includegraphics[clip,width=0.5\textwidth]{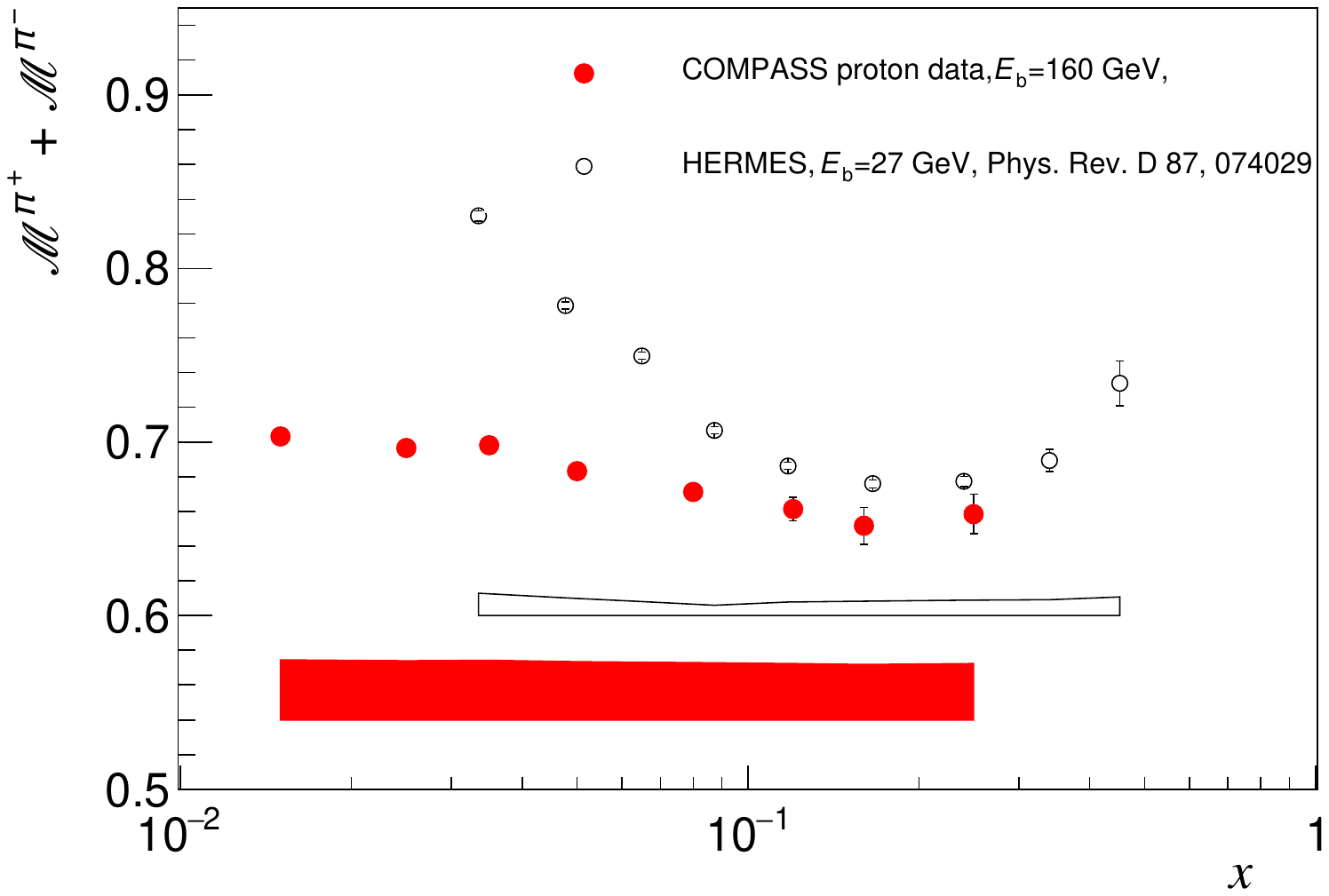}}
    \caption{Left panel: Sum of $\mathcal{M}^{\pi^+}\!\!+\!\mathcal{M}^{\pi^-}$ versus $x$ (for clarity staggered horizontally).
    Right panel: Comparison of $\mathcal{M}^{\pi^+}\!\!+\!\mathcal{M}^{\pi^-}$ versus $x$ for
    COMPASS and HERMES experiments. See text for details.}
    \label{fig:res-4}
\end{figure}

The left panel of Fig.~\ref{fig:res-4} compares $\mathcal{M}^{\pi^+}\!\!+\! \mathcal{M}^{\pi^-}$ for the present data on the proton target with the results on an isoscalar target~\cite{comp_pi}. Additionally, the
results of the present analysis using the same radiative correction procedure as in
Ref.~\cite{comp_pi} (previous RC) are presented.
In the case of the two latter results,
the data points are shown with total uncertainties.
As expected in pQCD, only a weak $x$-dependence is observed
in all the COMPASS multiplicity results.
Within the depicted total uncertainties,
the results published in Ref.~\cite{comp_pi} and those from the present analysis
with previous RC agree within one standard deviation. This takes into account the correlation across different bins, as discussed in Section \ref{sec:an:sys}.
In the right panel of Fig.~\ref{fig:res-4}, a comparison of the sum of
$\mathcal{M}^{\pi^+}\!\!+\!\mathcal{M}^{\pi^-}$ versus $x$ is shown for the present COMPASS
analysis of 160 GeV $\mu^{\pm}$p interactions and the HERMES \cite{hermes} 27.5 GeV $e^{\pm}$p interactions for the so called $x$-$z$ representation of their data. The COMPASS results
exhibit significant differences compared to the HERMES ones, although smaller than those in Ref.~\cite{comp_pi}.
With increasing energy, more particles of various types per interaction are produced in COMPASS than in HERMES. This means that the average $\langle z \rangle$ of hadrons in COMPASS is lower than in HERMES.
Thus, it is expected that in COMPASS $\mathcal{M}^{\pi^+}\!\!+\!\mathcal{M}^{\pi^-}$ is lower, as the integration over $z$ starts at $z=0.2$. While this may explain the lower values,
it does not explain the difference in shape\footnote{It is worth noting that the unusual shape of the HERMES data as a function of $x$ can be represented as a straight line as a function of $\langle y \rangle$.
The observed slope can be influenced by radiative corrections. Since the values of RC in Ref.~\cite{hermes} are not provided, we cannot compare them to those we can currently obtain using DJANGOH-MC for HERMES kinematics.}.

\begin{figure}
    \centering
    \includegraphics[clip,width=1.0\textwidth]{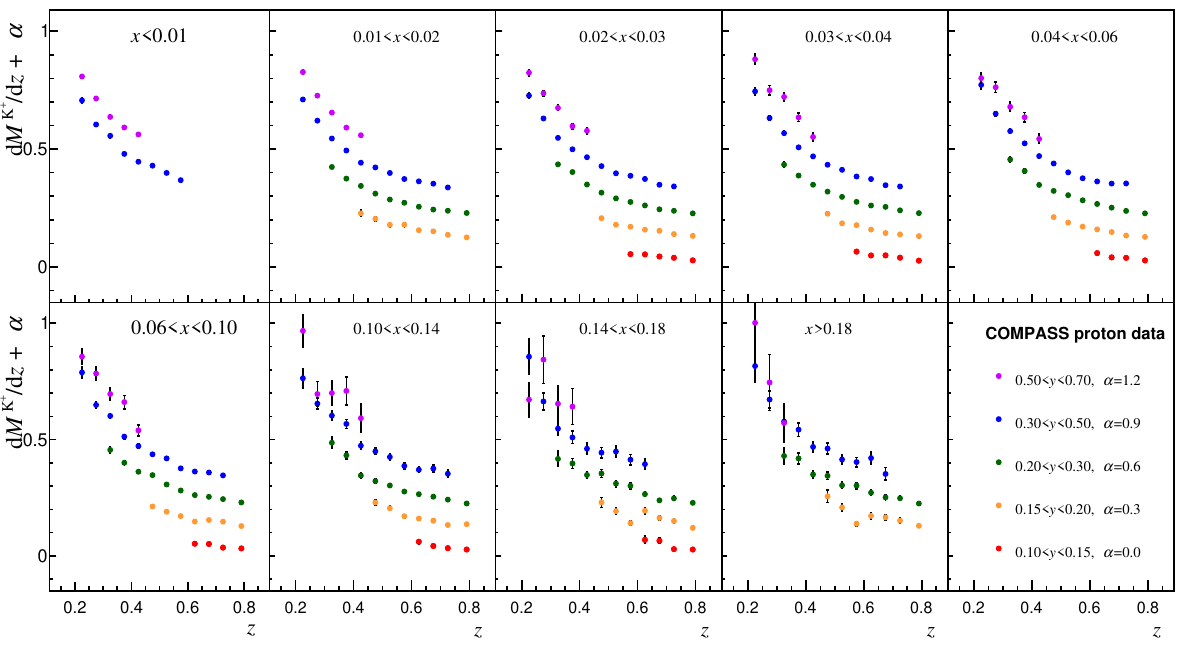}
    \caption{Positive kaon multiplicities versus $z$ for nine $x$ bins and five $y$ bins (for clarity staggered vertically by $\alpha$). Only statistical uncertainties are shown.}
    \label{fig:res-5}
\end{figure}

\begin{figure}
    \centering
    \includegraphics[clip,width=1.0\textwidth]{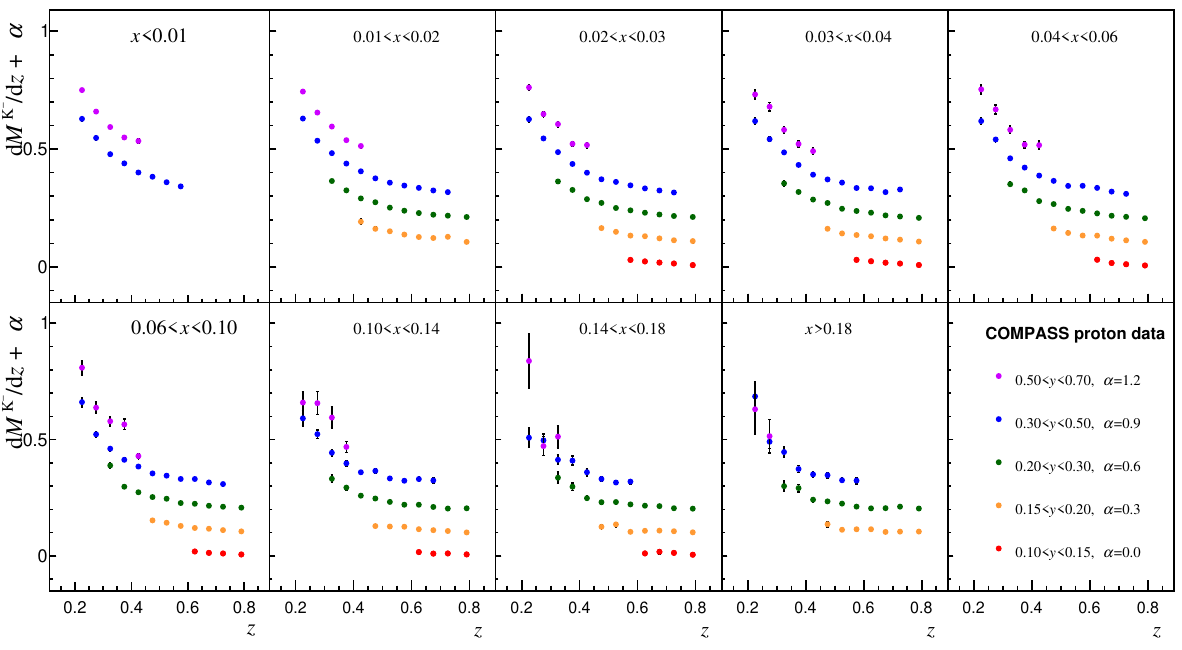}
    \caption{Same as Fig. \ref{fig:res-6} for negative kaons.}
    \label{fig:res-6}
\end{figure}

\begin{figure}
    \centering
    \includegraphics[clip,width=1.0\textwidth]{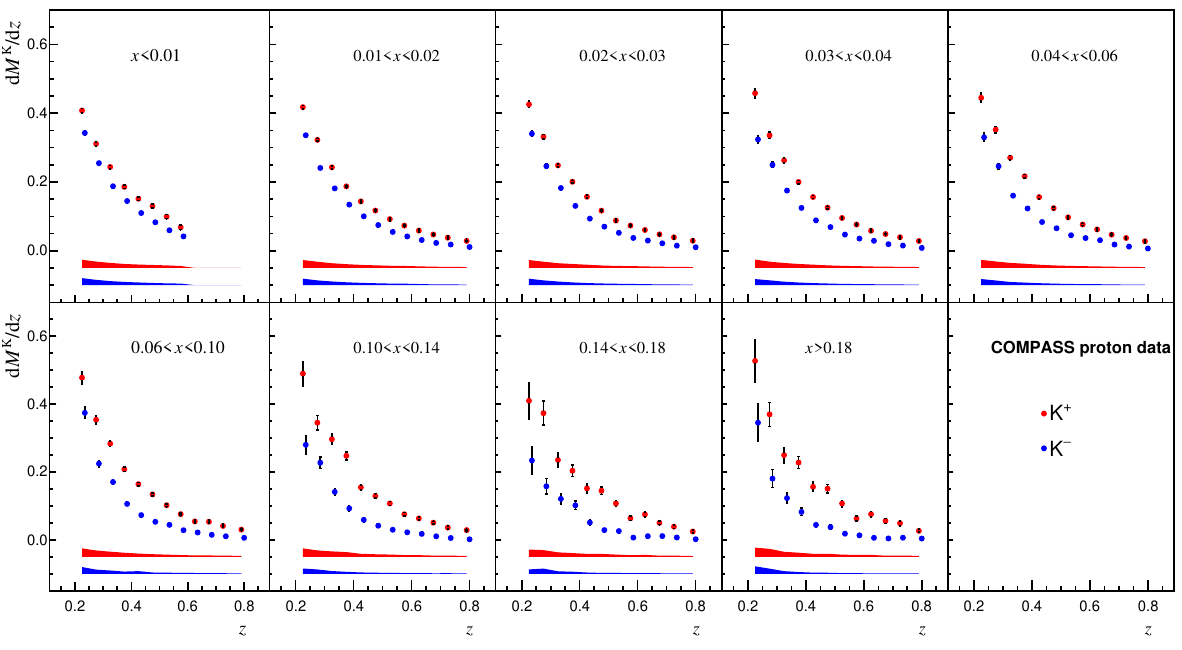}
    \caption{Positive (red) and negative (blue) kaon multiplicities versus $z$ (for clarity staggered horizontally) for nine $x$ bins.
 The data points are shown with statistical uncertainties, while the bands indicate systematic uncertainties.}
    \label{fig:res-65}
\end{figure}

In Figs \ref{fig:res-5} and \ref{fig:res-6}, we present multiplicity results for kaons of both electric charges, in the same manner as for unidentified hadrons
in Figs \ref{fig:res-1} and \ref{fig:res-2}.
The multiplicities for kaons are approximately five times lower than those for pions and a larger difference between positive and negative kaons is observed at high $x$. This difference is easily noticeable when comparing the results in Fig.~\ref{fig:res-65}, where $y$-averaged kaon results are presented, to those of pions in Fig.~\ref{fig:res-3}. This is attributed to the dominance of u-quarks and the fact that the valence d-quark in the proton target is also a valence quark of negative pions but not of negative kaons.

\begin{figure}
    \centerline{
    \includegraphics[clip,width=0.5\textwidth]{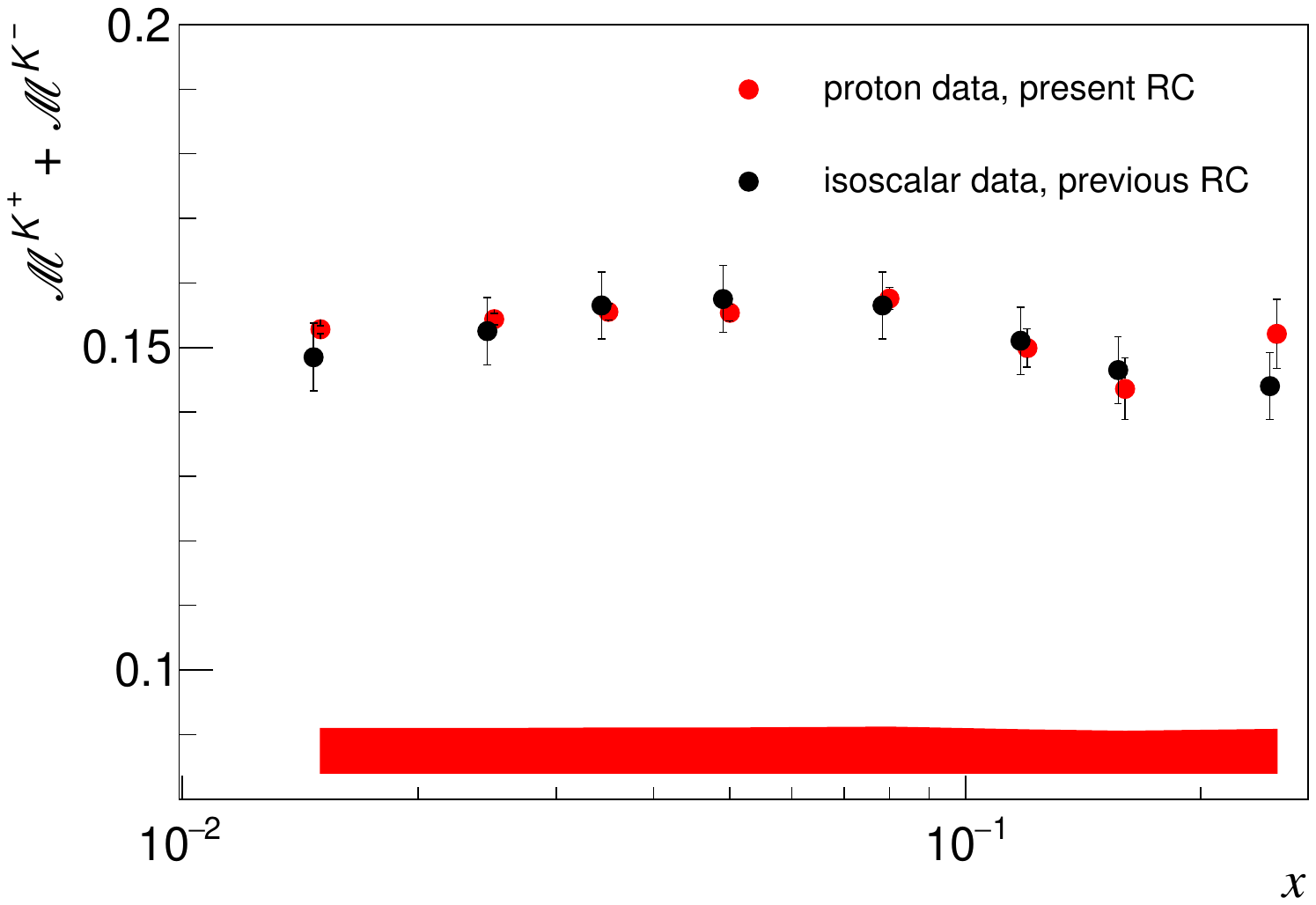}     \includegraphics[clip,width=0.5\textwidth]{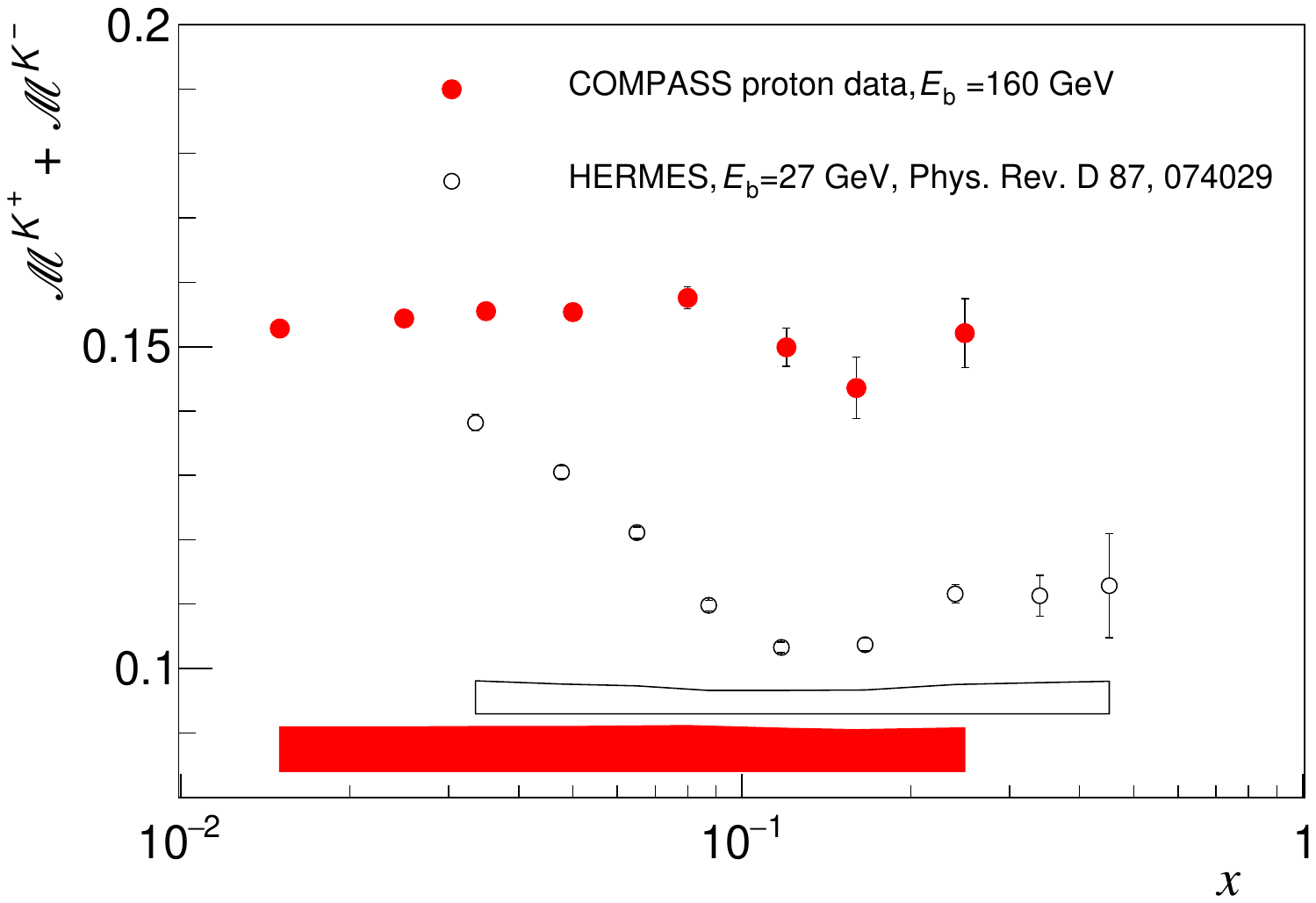}}
        \caption{Left panel: Sum of $\mathcal{M}^{K^+}\!\!+\!\mathcal{M}^{K^-}$ versus $x$ (for clarity staggered horizontally).
    Right panel: Comparison of $\mathcal{M}^{K^+}\!\!+\!\mathcal{M}^{K^-}$ versus $x$ for
    COMPASS and HERMES experiments. See text for details. }
    \label{fig:res-7}
\end{figure}

In the left panel of Fig.\ref{fig:res-7}, we provide a comparison of
$\mathcal{M}^{\rm K^+}\!\!+\!\mathcal{M}^{\rm K^-}$ as a function of $x$ for the present analysis and for the results in Ref.~\cite{comp_K}. For the isoscalar analysis, the total uncertainty is shown,
while for the present analysis, statistical and systematic uncertainties are separated.
A very good agreement between the two sets is observed. This is mostly due to the use of a more conservative estimate of RC in Ref.~\cite{comp_K} compared to
Ref.~\cite{comp_pi}, as shown in Fig.\ref{fig:rc2}. Particularly in the low $z$ region,
which yields the dominant contribution to
$\mathcal{M}^{\rm K^+}\!\!+\!\mathcal{M}^{\rm K^-}$, the present RC and those used in Ref. \cite{comp_K} are similar.
We observe a weak $x$-dependence of $\mathcal{M}^{\rm K^+}\!\!+\!\mathcal{M}^{\rm K^-}$, which might help to investigate
the role of strange quarks in pQCD.

In the right panel of Fig.~\ref{fig:res-7}, we compare $\mathcal{M}^{\rm K^+}\!\!+\! \mathcal{M}^{\rm K^-}$
as a function of $x$ for the present analysis and the HERMES data \cite{hermes}.
Notably, the COMPASS results are higher than the HERMES results, especially at larger values of $x$.
It is important to mention that the COMPASS results for the multiplicity ratio of
negative to positive kaons at high $z$ fall below the lower limit expected from pQCD ~\cite{comp_rk}.
This strong disagreement is more
pronounced at the lower centre-of-mass energy of the $\gamma^{*}$p system. Given that the
energy in the centre-of-mass for any $z$ in HERMES is lower than in COMPASS, the observed
difference may have a physics origin.

\begin{figure}
    \centering
    \includegraphics[clip,width=1.0\textwidth]{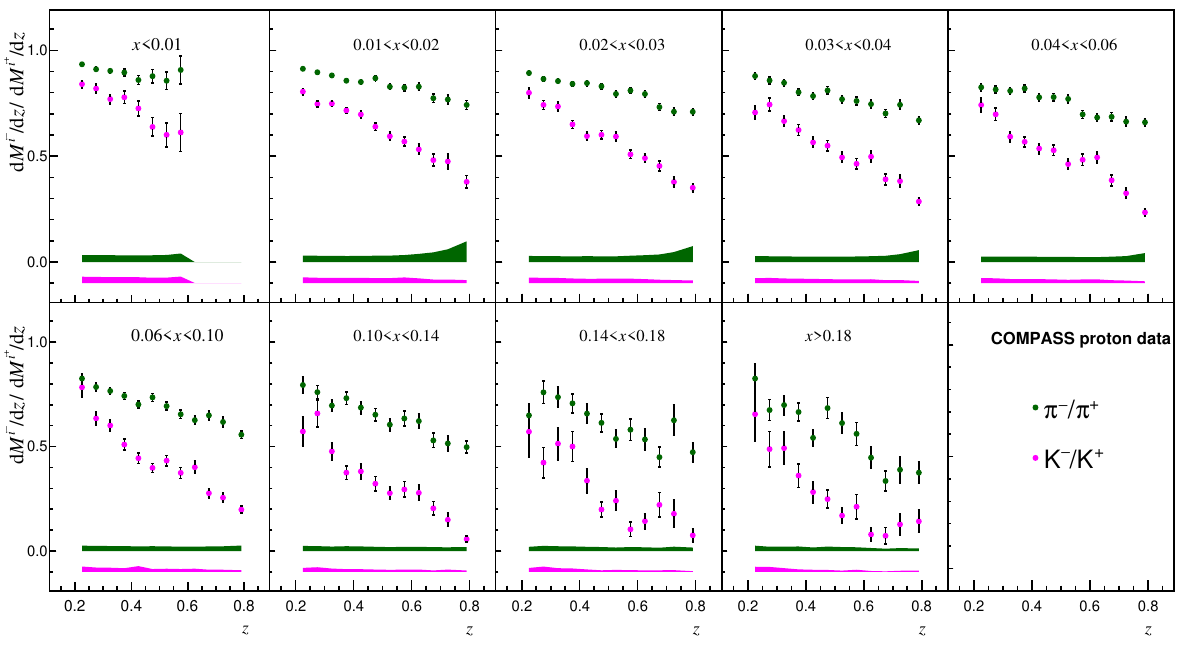}
    \caption{The multiplicity ratios of $\pi^{-}$ over $\pi^{+}$ and  K$^{-}$ over K$^{+}$ as a function
    of $z$ for nine bins of $x$. The $y$-averaged data were used, which for pions corresponds to the one presented in Fig.~\ref{fig:res-3}
    \label{fig:res-8}. The data points are shown with statistical uncertainties, while the bands indicate systematic uncertainties.}
\end{figure}

In Fig.~\ref{fig:res-8}, we present the ratio of K$^{-}$ to K$^{+}$ multiplicities as a
function of $z$ in nine bins of $x$ for data averaged over $y$. In this multiplicity ratio,
all correlated systematic effects cancel,
resulting in reduced relative systematic uncertainties
compared to standard multiplicities.
A clear and steep downward slope as a function of $z$ is evident. While the analysis of data in the
high-$z$ region is beyond the scope of the present paper, within the observed region
there appears to be no contradiction with the conclusions reached in Refs~\cite{comp_rk, comp_rkp} that for high-$z$ in the COMPASS kinematics the lower limit of (N)LO pQCD is violated.
For completeness, the results for the ratio of $\pi^{-}$ to $\pi^{+}$ multiplicities are also
presented in Fig.~\ref{fig:res-8}.
For pion data, the observed $z$ dependence appears to be weaker compared to that of kaons.
It is worth mentioning that, especially for high-$z$ and low-$x$ values,
the systematic uncertainties are relatively large,
mostly due to large uncertainties associated with the VM correction in this region.

\section{Summary}

Differential multiplicities of charge-separated pions, kaons and unidentified hadrons in
deep inelastic scattering of muons on a proton target
are presented.
They are provided in three-dimensional bins of $x$, $y$ and $z$,
covering the kinematic range $Q^2 > 1$ (GeV/$c)^2$, $0.004 < x < 0.4$ and $0.2 < z < 0.85$.
The numerical results of these measurements are available in HEPData, both with and without the
subtraction of the contribution from diffractive vector-meson production to SIDIS.
Furthermore, radiative correction factors are also given, as they were for the first time determined in COMPASS using the DJANGOH
Monte-Carlo generator.
Our high-precision, multi-dimensional data provides valuable input for future pQCD fits of fragmentation functions,
complementing earlier COMPASS isoscalar measurements.
The update of the published results on isoscalar targets with new radiative corrections is in preparation.

Considering the different radiative corrections, the results of the present analysis and the earlier
isoscalar measurements are in good agreement.
A significant discrepancy between the results of COMPASS
(using a 160 GeV muon beam) and HERMES (using a 27.5 GeV electron beam) is observed,
 particularly for the sum of K$^+$ and K$^-$ multiplicities when averaged over $y$ and integrated over $z$.
The observed discrepancy may have an origin beyond the scope of perturbative
quantum chromodynamics. Consequently, it becomes crucial to conduct similar measurements at
the Jefferson laboratory using a 12 GeV electron beam and at the future Electron-Ion Collider.

\section*{Acknowledgements}
We express our gratitude to H. Spiesberger for his valuable assistance in the use of the DJANGOH generator, which significantly contributed to its successful
implementation for COMPASS. We gratefully acknowledge the support of the CERN management and staff and the skill and effort of
the technicians of our collaborating institutes. This work was made possible by the financial support of
our funding agencies.

\clearpage

\center{\textbf{The COMPASS Collaboration}}

\vspace{10pt}
\begin{flushleft}
G.~D.~Alexeev$^\textrm{{\footnotesize\hyperlink{hl:dubna}{28}}}$\orcidlink{0009-0007-0196-8178},
M.~G.~Alexeev$^\textrm{{\footnotesize\hyperlink{hl:turin_u}{20},\hyperlink{hl:turin_i}{19}}}$\orcidlink{0000-0002-7306-8255},
C.~Alice$^\textrm{{\footnotesize\hyperlink{hl:turin_u}{20},\hyperlink{hl:turin_i}{19}}}$\orcidlink{0000-0001-6297-9857},
A.~Amoroso$^\textrm{{\footnotesize\hyperlink{hl:turin_u}{20},\hyperlink{hl:turin_i}{19}}}$\orcidlink{0000-0002-3095-8610},
V.~Andrieux$^\textrm{{\footnotesize\hyperlink{hl:illinois}{33}}}$\orcidlink{0000-0001-9957-9910},
V.~Anosov$^\textrm{{\footnotesize\hyperlink{hl:dubna}{28}}}$\orcidlink{0009-0003-3595-9561},
K.~Augsten$^\textrm{{\footnotesize\hyperlink{hl:praguectu}{4}}}$\orcidlink{0000-0001-8324-0576},
W.~Augustyniak$^\textrm{{\footnotesize\hyperlink{hl:warsaw}{23}}}$,
C.~D.~R.~Azevedo$^\textrm{{\footnotesize\hyperlink{hl:aveiro}{26}}}$\orcidlink{0000-0002-0012-9918},
B.~Badelek$^\textrm{{\footnotesize\hyperlink{hl:warsawu}{25}}}$\orcidlink{0000-0002-4082-1466},
J.~Barth$^\textrm{{\footnotesize\hyperlink{hl:bonniskp}{8}}}$\orcidlink{0009-0003-0891-9935},
R.~Beck$^\textrm{{\footnotesize\hyperlink{hl:bonniskp}{8}}}$,
J.~Beckers$^\textrm{{\footnotesize\hyperlink{hl:munichtu}{12}}}$\orcidlink{0009-0009-7186-255X},
Y.~Bedfer$^\textrm{{\footnotesize\hyperlink{hl:saclay}{6}}}$\orcidlink{0000-0002-5198-1852},
J.~Bernhard$^\textrm{{\footnotesize\hyperlink{hl:cern}{30}}}$\orcidlink{0000-0001-9256-971X},
M.~Bodlak$^\textrm{{\footnotesize\hyperlink{hl:praguecu}{5}}}$,
F.~Bradamante$^\textrm{{\footnotesize\hyperlink{hl:triest_i}{17}}}$\orcidlink{0000-0001-6136-376X},
A.~Bressan$^\textrm{{\footnotesize\hyperlink{hl:triest_u}{18},\hyperlink{hl:triest_i}{17}}}$\orcidlink{0000-0002-3718-6377},
W.-C.~Chang$^\textrm{{\footnotesize\hyperlink{hl:taipei}{31}}}$\orcidlink{0000-0002-1695-7830},
C.~Chatterjee$^\textrm{{\footnotesize\hyperlink{hl:triest_i}{17},\hyperlink{hl:b}{b}}}$\orcidlink{0000-0001-7784-3792},
M.~Chiosso$^\textrm{{\footnotesize\hyperlink{hl:turin_u}{20},\hyperlink{hl:turin_i}{19}}}$\orcidlink{0000-0001-6994-8551},
S.-U.~Chung$^\textrm{{\footnotesize\hyperlink{hl:munichtu}{12},\hyperlink{hl:j}{j},\hyperlink{hl:j1}{j1}}}$,
A.~Cicuttin$^\textrm{{\footnotesize\hyperlink{hl:triest_i}{17},\hyperlink{hl:triest_a}{16}}}$\orcidlink{0000-0002-3645-9791},
P.~M.~M.~Correia$^\textrm{{\footnotesize\hyperlink{hl:aveiro}{26}}}$\orcidlink{0000-0001-7292-7735},
M.~L.~Crespo$^\textrm{{\footnotesize\hyperlink{hl:triest_i}{17},\hyperlink{hl:triest_a}{16}}}$\orcidlink{0000-0002-5483-3388}
% ...

\vspace{10pt}
\hypertarget{hl:aanl}{$^\textrm{{\footnotesize 1}}$\footnotesize~A.I. Alikhanyan National Science Laboratory, 2 Alikhanyan Br. Street, 0036, Yerevan, Armenia$^\textrm{{\tiny\hyperlink{hl:A}{A}}}$\\}
\hypertarget{hl:brno}{$^\textrm{{\footnotesize 2}}$\footnotesize~Institute of Scientific Instruments of the CAS, 61264 Brno, Czech Republic$^\textrm{{\tiny\hyperlink{hl:B}{B}}}$\\}
\hypertarget{hl:liberec}{$^\textrm{{\footnotesize 3}}$\footnotesize~Technical University in Liberec, 46117 Liberec, Czech Republic$^\textrm{{\tiny\hyperlink{hl:B}{B}}}$\\}
\hypertarget{hl:praguectu}{$^\textrm{{\footnotesize 4}}$\footnotesize~Czech Technical University in Prague, 16636 Prague, Czech Republic$^\textrm{{\tiny\hyperlink{hl:B}{B}}}$\\}
\hypertarget{hl:praguecu}{$^\textrm{{\footnotesize 5}}$\footnotesize~Charles University, Faculty of Mathematics and Physics, 12116 Prague, Czech Republic$^\textrm{{\tiny\hyperlink{hl:B}{B}}}$\\}
\hypertarget{hl:saclay}{$^\textrm{{\footnotesize 6}}$\footnotesize~IRFU, CEA, Universit\'e Paris-Saclay, 91191 Gif-sur-Yvette, France\\}
\hypertarget{hl:bochum}{$^\textrm{{\footnotesize 7}}$\footnotesize~Universit\"at Bochum, Institut f\"ur Experimentalphysik, 44780 Bochum, Germany$^\textrm{{\tiny\hyperlink{hl:C}{C}}}$\\}
\hypertarget{hl:bonniskp}{$^\textrm{{\footnotesize 8}}$\footnotesize~Universit\"at Bonn, Helmholtz-Institut f\"ur  Strahlen- und Kernphysik, 53115 Bonn, Germany$^\textrm{{\tiny\hyperlink{hl:C}{C}}}$\\}
\hypertarget{hl:bonnpi}{$^\textrm{{\footnotesize 9}}$\footnotesize~Universit\"at Bonn, Physikalisches Institut, 53115 Bonn, Germany$^\textrm{{\tiny\hyperlink{hl:C}{C}}}$\\}
\hypertarget{hl:freiburg}{$^\textrm{{\footnotesize 10}}$\footnotesize~Universit\"at Freiburg, Physikalisches Institut, 79104 Freiburg, Germany$^\textrm{{\tiny\hyperlink{hl:C}{C}}}$\\}
\hypertarget{hl:mainz}{$^\textrm{{\footnotesize 11}}$\footnotesize~Universit\"at Mainz, Institut f\"ur Kernphysik, 55099 Mainz, Germany$^\textrm{{\tiny\hyperlink{hl:C}{C}}}$\\}
\hypertarget{hl:munichtu}{$^\textrm{{\footnotesize 12}}$\footnotesize~Technische Universit\"at M\"unchen, Physik Dept., 85748 Garching, Germany$^\textrm{{\tiny\hyperlink{hl:C}{C}}}$\\}
\hypertarget{hl:munichuni}{$^\textrm{{\footnotesize 13}}$\footnotesize~Ludwig-Maximilians-Universit\"at, 80539 M\"unchen, Germany\\}
\hypertarget{hl:calcutta}{$^\textrm{{\footnotesize 14}}$\footnotesize~Matrivani Institute of Experimental Research \& Education, Calcutta-700 030, India$^\textrm{{\tiny\hyperlink{hl:D}{D}}}$\\}
\hypertarget{hl:telaviv}{$^\textrm{{\footnotesize 15}}$\footnotesize~Tel Aviv University, School of Physics and Astronomy, 69978 Tel Aviv, Israel$^\textrm{{\tiny\hyperlink{hl:E}{E}}}$\\}
\hypertarget{hl:triest_a}{$^\textrm{{\footnotesize 16}}$\footnotesize~Abdus Salam ICTP, 34151 Trieste, Italy\\}
\hypertarget{hl:triest_i}{$^\textrm{{\footnotesize 17}}$\footnotesize~Trieste Section of INFN, 34127 Trieste, Italy\\}
\hypertarget{hl:triest_u}{$^\textrm{{\footnotesize 18}}$\footnotesize~University of Trieste, Dept.\ of Physics, 34127 Trieste, Italy\\}
\hypertarget{hl:turin_i}{$^\textrm{{\footnotesize 19}}$\footnotesize~Torino Section of INFN, 10125 Torino, Italy\\}
\hypertarget{hl:turin_u}{$^\textrm{{\footnotesize 20}}$\footnotesize~University of Torino, Dept.\ of Physics, 10125 Torino, Italy\\}
\hypertarget{hl:miyazaki}{$^\textrm{{\footnotesize 21}}$\footnotesize~University of Miyazaki, Miyazaki 889-2192, Japan$^\textrm{{\tiny\hyperlink{hl:F}{F}}}$\\}
\hypertarget{hl:yamagata}{$^\textrm{{\footnotesize 22}}$\footnotesize~Yamagata University, Yamagata 992-8510, Japan$^\textrm{{\tiny\hyperlink{hl:F}{F}}}$\\}
\hypertarget{hl:warsaw}{$^\textrm{{\footnotesize 23}}$\footnotesize~National Centre for Nuclear Research, 02-093 Warsaw, Poland$^\textrm{{\tiny\hyperlink{hl:G}{G}}}$\\}
\hypertarget{hl:warsawtu}{$^\textrm{{\footnotesize 24}}$\footnotesize~Warsaw University of Technology, Institute of Radioelectronics, 00-665 Warsaw, Poland$^\textrm{{\tiny\hyperlink{hl:G}{G}}}$\\}
\hypertarget{hl:warsawu}{$^\textrm{{\footnotesize 25}}$\footnotesize~University of Warsaw, Faculty of Physics, 02-093 Warsaw, Poland$^\textrm{{\tiny\hyperlink{hl:G}{G}}}$\\}
\hypertarget{hl:aveiro}{$^\textrm{{\footnotesize 26}}$\footnotesize~University of Aveiro, I3N, Dept. of Physics, 3810-193 Aveiro, Portugal$^\textrm{{\tiny\hyperlink{hl:H}{H}}}$\\}
\hypertarget{hl:lisbon}{$^\textrm{{\footnotesize 27}}$\footnotesize~LIP, 1649-003 Lisbon, Portugal$^\textrm{{\tiny\hyperlink{hl:H}{H}}}$\\}
\hypertarget{hl:dubna}{$^\textrm{{\footnotesize 28}}$\footnotesize~Affiliated with an international laboratory covered by a cooperation agreement with CERN\\}
\hypertarget{hl:russia}{$^\textrm{{\footnotesize 29}}$\footnotesize~Affiliated with an institute covered by a cooperation agreement with CERN.\\}
\hypertarget{hl:cern}{$^\textrm{{\footnotesize 30}}$\footnotesize~CERN, 1211 Geneva 23, Switzerland\\}
\hypertarget{hl:taipei}{$^\textrm{{\footnotesize 31}}$\footnotesize~Academia Sinica, Institute of Physics, Taipei 11529, Taiwan$^\textrm{{\tiny\hyperlink{hl:I}{I}}}$\\}
\hypertarget{hl:taipeincu}{$^\textrm{{\footnotesize 32}}$\footnotesize~Center for High Energy and High Field Physics and Dept.\ of Physics, National Central University, 300 Zhongda Rd., Zhongli 320317, Taiwan$^\textrm{{\tiny\hyperlink{hl:I}{I}}}$\\}
\hypertarget{hl:illinois}{$^\textrm{{\footnotesize 33}}$\footnotesize~University of Illinois at Urbana-Champaign, Dept.\ of Physics, Urbana, IL 61801-3080, USA$^\textrm{{\tiny\hyperlink{hl:J}{J}}}$\\}

\vspace{10pt}
\hypertarget{hl:*}{$^\textrm{{\footnotesize *}}$\footnotesize~Corresponding author\\}
\hypertarget{hl:a}{$^\textrm{{\footnotesize a}}$\footnotesize~Supported by the Higher Education and Science Committee of Republic of Armenia, in the frame of the research project No 21AG-1C028, Armenia\\}
\hypertarget{hl:b}{$^\textrm{{\footnotesize b}}$\footnotesize~Supported by the European Union’s Horizon 2020 research and innovation programme under grant agreement STRONG–2020 - No 824093\\}
\hypertarget{hl:c}{$^\textrm{{\footnotesize c}}$\footnotesize~Retired from Ludwig-Maximilians-Universit\"at, 80539 M\"unchen, Germany\\}
\hypertarget{hl:c1}{$^\textrm{{\footnotesize c1}}$\footnotesize~Supported by the DFG cluster of excellence `Origin and Structure of the Universe' (www.universe-cluster.de) (Germany)\\}
\hypertarget{hl:d}{$^\textrm{{\footnotesize d}}$\footnotesize~Also at ORIGINS Excellence Cluster, 85748 Garching, Germany\\}
\hypertarget{hl:e}{$^\textrm{{\footnotesize e}}$\footnotesize~Also at Institut f\"ur Theoretische Physik, Universit\"at T\"ubingen, 72076 T\"ubingen, Germany\\}
\hypertarget{hl:f}{$^\textrm{{\footnotesize f}}$\footnotesize~Present address: NISER, Centre for Medical and Radiation Physics, Bubaneswar, India\\}
\hypertarget{hl:g}{$^\textrm{{\footnotesize g}}$\footnotesize~Also at University of Eastern Piedmont, 15100 Alessandria, Italy\\}
\hypertarget{hl:g1}{$^\textrm{{\footnotesize g1}}$\footnotesize~Supported by the Funds for Research 2019-22 of the University of Eastern Piedmont\\}
\hypertarget{hl:h}{$^\textrm{{\footnotesize h}}$\footnotesize~Also at Chubu University, Kasugai, Aichi 487-8501, Japan\\}
\hypertarget{hl:i}{$^\textrm{{\footnotesize i}}$\footnotesize~Also at KEK, 1-1 Oho, Tsukuba, Ibaraki 305-0801, Japan\\}
\hypertarget{hl:j}{$^\textrm{{\footnotesize j}}$\footnotesize~Also at Dept.\ of Physics, Pusan National University, Busan 609-735, Republic of Korea\\}
\hypertarget{hl:j1}{$^\textrm{{\footnotesize j1}}$\footnotesize~Also at Physics Dept., Brookhaven National Laboratory, Upton, NY 11973, USA\\}
\hypertarget{hl:k}{$^\textrm{{\footnotesize k}}$\footnotesize~Also at Fairmont State University, Department of Natural Sciences, 1201 Locust Ave, Fairmont, West Virginia 26554, USA\\}
\hypertarget{hl:l}{$^\textrm{{\footnotesize l}}$\footnotesize~Also at Dept.\ of Physics, National Kaohsiung Normal University, Kaohsiung County 824, Taiwan\\}
\hypertarget{hl:$\dagger$}{$^\textrm{{\footnotesize $\dagger$}}$\footnotesize~Deceased\\}

\vspace{10pt}
\hypertarget{hl:A}{$^\textrm{{\tiny A}}$\footnotesize~Supported by the Higher Education and Science Committee of the Republic of Armenia (Armenia)\\}
\hypertarget{hl:B}{$^\textrm{{\tiny B}}$\footnotesize~Supported by MEYS Grants LM2018104, LM2023040 and LTT17018 and Charles University grants PRIMUS/22/SCI/017 and GAUK60121 (Czech Republic)\\}
\hypertarget{hl:C}{$^\textrm{{\tiny C}}$\footnotesize~Supported by BMBF - Bundesministerium f\"ur Bildung und Forschung (Germany)\\}
\hypertarget{hl:D}{$^\textrm{{\tiny D}}$\footnotesize~Supported by B. Sen fund (India)\\}
\hypertarget{hl:E}{$^\textrm{{\tiny E}}$\footnotesize~Supported by the Israel Academy of Sciences and Humanities (Israel)\\}
\hypertarget{hl:F}{$^\textrm{{\tiny F}}$\footnotesize~Supported by MEXT and JSPS, Grants 18002006, 20540299, 18540281 and 26247032, the Daiko and Yamada Foundations (Japan)\\}
\hypertarget{hl:G}{$^\textrm{{\tiny G}}$\footnotesize~Supported by NCN, Grant 2020/37/B/ST2/01547 (Poland)\\}
\hypertarget{hl:H}{$^\textrm{{\tiny H}}$\footnotesize~Supported by FCT, Grants DOI 10.54499/CERN/FIS-PAR/0022/2019 and DOI 10.54499/CERN/FIS-PAR/0016/2021 (Portugal)\\}
\hypertarget{hl:I}{$^\textrm{{\tiny I}}$\footnotesize~Supported by the Ministry of Science and Technology (Taiwan)\\}
\hypertarget{hl:J}{$^\textrm{{\tiny J}}$\footnotesize~Supported by the National Science Foundation, Grant no. PHY-1506416 (USA)\\}

\end{flushleft}


\begin{thebibliography}{99}


 \bibitem{lep} ALEPH Collaboration, R. Barate, {\it et al.}, Phys. Rep. {\bf 294} (1998) 1; \newline
DELPHI Collaboration, P. Abreu, {\it et al.}, Eur. Phys. J. C {\bf 5} (1998) 585;  \newline
OPAL Collaboration, R. Akers, {\it et al.}, Z. Phys. C {\bf 63} (1994) 181.
\bibitem{slac} SLD Collaboration, K. Abe, {\it et al.}, Phys. Rev. D {\bf 69} (2004) 072003 \newline
BABAR Collaboration, J.P. Lees, {\it et al.}, Phys. Rev. D {\bf 88} (2013) 032011.
\bibitem{belle} BELLE Collaboration, M. Leitgab, {\it et al.}, Phys. Rev. Lett. {\bf 111} (2013) 062002.


\bibitem{hermes} HERMES Collaboration, A. Airapetian {\it et al.}, Phys. Rev. D {\bf 87} (2013) 074029.
\bibitem{emc} EMC, J. Ashman, {\it et al.}, Z. Phys. C {\bf 52} (1991) 361.
\bibitem{comp_pi} COMPASS Collaboration, C. Adolph {\it et al.}, Phys. Lett. B {\bf 764} (2017) 1.
\bibitem{comp_K} COMPASS Collaboration, C. Adolph {\it et al.}, Phys. Lett. B {\bf 767} (2017) 133.


\bibitem{rhic} PHENIX Collaboration, S.S. Adler, {\it et al.}, Phys. Rev. Lett. {\bf 91} (2003) 241803; \newline
STAR Collaboration, J. Adams, {\it et al.}, Phys. Rev. Lett. {\bf 97} (2006) 152302; \newline\
BRAHMS Collaboration, I. Arsene, {\it et al.}, Phys. Rev. Lett. {\bf 98} (2007) 252001; \newline STAR Collaboration, B.I. Abelev, {\it et al.}, Phys. Rev. C {\bf 75} (2007) 064901.
\bibitem{dss_01} D. de Florian, R. Sassot and M. Stratmann, Phys. Rev. D {\bf 75} (2007) 114010.

\bibitem{hkns} M. Hirai, S. Kumano, T.-H. Nagai and K. Sudoh, Phys. Rev. D {\bf 75} (2007) 094009.

\bibitem{lss} E. Leader, A.V. Sidorov and D. Stamenov, Phys. Rev. D {\bf 93} (2016) 074026.

\bibitem{dss_02} D. de Florian, {\it et al.}, Phys. Rev. D {\bf 91} (2015) 014035.
\bibitem{dss_03} D. de Florian, {\it et al.}, Phys. Rev. D {\bf 95} (2017) 094019.

\bibitem{nnpdf_ff} V. Bertone {\it et al.},  Eur. Phys. J. C {\bf 77} (2017) 516.
\bibitem{jam_ff} E. Moffat, W. Melnitchouk, T. C. Rogers and N. Sato Phys. Rev. D {\bf 104}, (2021) 016015.
\bibitem{nnlo_ff} I. Borsa {\it et al.}, Phys. Rev. Lett. {\bf 129} (2022) 012002.


\bibitem{nlo_sidis} D. de Florian, M. Stratmann and W. Vogelsang, Phys. Rev. D {\bf 57} (1998) 5811.


\bibitem{comp_exp} P. Abbon, {\it et al.}, Nucl. Instrum. Meth. A {\bf 631} (2011) 26.




\bibitem{lepto} G. Ingelman, A. Edin and J. Rathsman, Comput.\ Phys.\ Commun.\  {\bf 101} (1997) 108.
\bibitem{lund} T. Sj\"ostrand, LU-TP-95-20, CERN-TH-7112-93-REV, hep-ph/9508391.

\bibitem{comp_dg} COMPASS Collaboration, C. Adolph, {\it et al.}, Phys. Lett. B {\bf 718} (2013) 922.

\bibitem{geant4} GEANT4 Collaboration, S. Agostinelli, {\it et al.}, Nucl. Inst. Meth. A {\bf 506} (2003) 250.

\bibitem{hepgen} A. Sandacz and P. Sznajder, arXiv:1207.0333.

\bibitem{hepgen_model} S.V. Goloskokov and P. Kroll, Eur. Phys. J. C {\bf 53} (2008) 367.

\bibitem{terad} A.A. Akhundov, D. Bardin, L. Kalinovskaya and T. Riemann, Fortschr. Phys. {\bf 44} (1996) 373.

\bibitem{djangoh} E. C. Aschenauer {\it et al.},  Phys. Rev. D {\bf 88} (2013) 114025.

\bibitem{hep_data} The Durham HEPData Project, http://durpdg.dur.ac.uk/.

\bibitem{comp_rk} COMPASS Collaboration, R. Akhunzyanov {\it et al.}, Phys. Lett. B {\bf 786} (2018) 390.
\bibitem{comp_rkp} COMPASS Collaboration, M.G. Alexeev {\it et al.},  Phys. Lett. B {\bf 807} (2020) 135600.




\end{thebibliography}
\end{document}